\documentstyle[aps,epsf,floats]{revtex}
\newcommand{\re}{\mathop{\mathrm{Re}}}
\newcommand{\im}{\mathop{\mathrm{Im}}}
\newcommand{\Tr}{\mathop{\mathrm{Tr}}}
\newcommand{\diag}{\mathop{\mathrm{diag}}}
\newcommand{\Vol}{\mathop{\mathrm{Vol}}}
\newcommand{\Str}{\mathop{\mathrm{Str}}}

\begin{document}
\author{
Yan V. Fyodorov$\S$\thanks{
On leave from Petersburg Nuclear Physics Institute, Gatchina
188350, Russia},
Boris A. Khoruzhenko$\P$ and Hans-J\"{u}rgen Sommers$\S$
       }
\address{
$\S$Fachbereich Physik, Universit\"at-GH Essen,
D-45117 Essen, Germany
        }
\address{
$\P$ School of Mathematical Sciences, Queen Mary \& Westfield
College, \\ University of London, London E1 4NS, U.K.
        }

\title{Universality in the random matrix spectra in the
regime of weak non-Hermiticity}

\date{15 January 1998}
\maketitle
\begin{abstract}
Complex spectra of random matrices are studied in the regime of
weak non-Hermiticity. The matrices we consider are of the form
$\hat{H}_1+i\hat{H}_2$, where $\hat{H}_1$ and $\hat{H}_2$
are Hermitian and statistically independent. In the first part of the paper we consider the case of matrices
$\hat H_{1,2}$ having i.i.d.\ entries.
For such matrices the regime of weak
non-Hermiticity is defined in the limit of large matrix dimension $N$
by the condition $\langle \Tr \hat{H}_1^2 \rangle
\propto N \langle \mbox{Tr} \hat{H}_2^2 \rangle $.
 We show that in the regime of weak non-Hermiticity
the distribution of complex eigenvalues of $\hat{H}_1+i\hat{H}_2$
 is dictated by the global symmetries of $\hat{H}_{1,2}$,
but otherwise is universal, i.e.\
independent of the particular distributions of their entries.
Our heuristic proof is based on
the supersymmetric technique and extends also to ``invariant''
ensembles of $\hat{H}_{1,2}$.
In the second part of the paper we study Gaussian complex matrices
in the regime of weak non-Hermiticity. Using the
mathematically rigorous method of orthogonal polynomials
we find the eigenvalue correlation functions. This
allows us to obtain explicitly various eigenvalue statistics.
These statistics describe a crossover from Hermitian matrices
characterized by the Wigner-Dyson
statistics of real eigenvalues to strongly non-Hermitian matrices
whose complex eigenvalues were studied by Ginibre.
Two-point statistical measures such as
spectral form-factor, number variance and
small distance behavior of the nearest neighbor
distance distribution $p(s)$ are studied thoroughly.
In particular, we found that the latter function
may exhibit unusual behavior
$p(s)\propto s^{5/2}$ for some parameter values.

\end{abstract}

\section{Introduction}

Eigenvalues of large random matrices
have been attracting much interest in theoretical
physics since the 1950's
\cite{Port,Mehta,Haake_book,Bohigas,AlSi,Bee,Guhr}.
Until recently only the real eigenvalues were seen as
physically relevant, hence most of the studies
ignored matrices with complex eigenvalues. Powerful techniques to deal
with real eigenvalues were developed and their statistical properties
are well understood nowadays \cite{Mehta}. Microscopic justifications of the use of random matrices
for describing the universal spectral properties of quantum chaotic systems
have been
provided by several groups recently, based both on traditional
semiclassical
periodic orbit expansions \cite{per,bogomol} and on advanced
field-theoretical methods \cite{MK,aaas}.
These facts make the theory of random Hermitian
matrices a powerful and versatile
tool of research in different branches of modern theoretical physics,
see e.g.\cite{Bohigas,Bee,Guhr}.

Recent studies of dissipative quantum maps \cite{diss,reichl},
asymmetric neural networks \cite{neural,n}, and open quantum
systems \cite{Sok,nils,dittes,FS,FSR} stimulated interest
to complex
eigenvalues of random matrices. Most obvious
motivation comes from studies of resonances in
open quantum systems,
i.e.\ systems whose fragments can escape to or come from
infinity. The resonances are determined as poles of the
scattering matrix (S-matrix), as a function of energy of incoming waves,
in the complex energy plane. The real part of the pole is the resonance
energy and the imaginary part is the resonance half-width. Finite width
implies
finite life-time of the corresponding states. In the chaotic regime
the resonances are dense and placed irregularly in the complex plane.
Recently, the progress in numerical techniques and
computational facilities made available  resonance patterns of
high accuracy for realistic open quantum chaotic systems
like atoms and molecules \cite{Blumel}.

Due to irregularity in the resonance widths and positions
the S-matrix shows
irregular fluctuations with energy and the main goal of the
theory of the chaotic scattering is to provide an adequate statistical
description of such a behavior.
The so-called ``Heidelberg approach''
to this problem suggested in \cite{VWZ} makes use of random matrices.
The starting point is
a representation of the S-matrix in terms of an
effective non-Hermitian Hamiltonian
${\cal H}_{eff}=\hat{H}-i\hat{\Gamma}$. The Hermitian $N\times N$ matrix
$\hat{H}$ describes the closed counterpart of the open
system and the skew-Hermitian $i\hat{\Gamma}=i\hat W \hat W^T$
arises due to coupling to open scattering channels $a=1, \ldots , M$,
the matrix elements $W_{ja}$ being the amplitudes of direct transitions
from "internal" states $i=1,\ldots , N$ to one of open channels.
The poles of the S-matrix coincide
with the eigenvalues of ${\cal H}_{eff}$. In the chaotic regime
one replaces
$\hat{H}$ with an ensemble of random matrices of an
appropriate symmetry. This step is usually
``justified'' by the common belief according
to which the universal features of the
chaotic quantum systems survive such a replacement \cite{Bohigas,AlSi,Bee,Guhr}.   As a result, various features of
chaotic quantum scattering
can be efficiently studied by performing the ensemble averaging.
The approach has proved to be very fruitful (for an account of
recent developments see \cite{FSR}). In particular,
it allowed to
obtain explicitly the distribution of the resonances
in the complex plane for chaotic quantum systems
with broken time-reversal invariance \cite{FS,FSR} and,
in its turn, this distribution was used to clarify some
aspects of the relaxation
processes in quantum chaotic systems\cite{Sav}.

A very recent outburst of interest to the non-Hermitian problems
\cite{Nels,NS,pass,Efnonh,freevs,nhloc,zee,NS1,QCD}
deserves to be mentioned separately. During the last several
years complex spectra of random matrices and operators
emerged in a diversity of problems. Hatano and Nelson
described depinning of flux lines from columnar defects
in superconductors in terms
of a localization-delocalization transition
in non-Hermitian quantum mechanics \cite{Nels}. Their work
motivated a series of studies of the
corresponding non-Hermitian
Schr\"odinger operator \cite{Efnonh,freevs,nhloc,zee,NS1}
and, surprisingly, random matrices appeared to be relevant
in this context \cite{Efnonh,freevs}. Complex eigenvalues
were also discussed in the context of lattice
QCD. The lattice Dirac operator entering the QCD
partition function is non-Hermitian at nonzero
chemical potential and proves to be difficult to deal with
both numerically and
analytically. Recent studies of chiral symmetry breaking used
a non-Hermitian random matrix substitute for the
Dirac operator
\cite{QCD}. There exist also interesting
links between complex eigenvalues of random matrices and
systems of interacting particles in one and two spatial dimensions
\cite{cal}. And, finally, we have to mention that
random matrices can be used for visualization of
the pseudospectra of non-random
convection-diffusion operators \cite{Tref}, and for description of two-level systems coupled to the noise reservoir\cite{ewa}.

Traditional mathematical treatment of random matrices
with no symmetry conditions imposed
goes back to the pioneering work by Ginibre\cite{Gin}
who determined all the eigenvalue correlation functions
in an ensemble of complex matrices with Gaussian entries.
The progress in the field was rather slow but steady
\cite{Mehta,Lehm1,Edel,Forr,Oas,FKS2}, see also \cite{Gir,Bai}.
In addition to the traditional approach
other aproaches have been developed and tested on new classes of
non-Hermitian random matrices
\cite{n,nils,FS,Khor,FKS1,free,zee1,Kus}.
However, our knowledge of the statistical properties of
complex eigenvalues of random matrices is still far from being
complete,
in particular little is known about the universality classes of
the obtained eigenvalue statistics.

When speaking about universality one has to specify
the energy scale, for the degree of universality
depends usually upon the chosen scale.
There exist two characteristic scales in the random matrix
spectra: the global one and the local one.
The global scale is aimed at description of the distribution of
the eigenvalues in bulk. The local one is aimed at decription of the
statistical properties  of small eigenvalue sets. For real spectra,
the global scale is that on which
a spectral interval of
unit length contains on average a large, proportional to
the matrix dimension, number of eigenvalues. If
the spectrum is supported in a finite interval $[a,b]$
the global scale is simply given by the length of this interval.
In contrary, the local scale
is that
determined by the mean distance $\Delta $ between two
neighbouring eigenvalues.
Loosely speaking, the local scale is $N$ times smaller than
the global one sufficiently far from the spectrum edges,
$N$ being the matrix dimension.

Universality in the real spectra is well
established. The global scale universality is specific
to random matrices with independent entries
and does not extend to other classes of random matrices.
The best known example of such universality is provided
by the Wigner semicircle law \cite{Wig}:
\begin{equation}\label{semi}
\langle\rho(X)\rangle=\frac{N}{2\pi J^2 }
\sqrt{4\,J^2-X^2}=N\nu(X)=\frac{1}{\Delta},
\end{equation}
 which holds
for random matrices whose entries satisfy a Lindeberg type condition
\cite{pastur}.  In this expression the parameter $J$ just sets the global scale
in a sense as defined above. It is determined by the expectation value
$J^2=\langle \frac{1}{N}\Tr\hat{H}^2\rangle$.
It is generally accepted to scale
 entries in such a way that $J$ stays finite when $N\to \infty$,
the local spacing between eigenvalues in the neighbourhood of the
point $X$ being therefore $\Delta\propto 1/N$.
Similar universality is also known for
complex spectra \cite{Gir,Bai}.

  From the point of view of
universality the semicircular eigenvalue density
is not extremely robust.
Most easily one violates it by considering an
important class of so-called "invariant ensembles"
characterized by a probability density
of the form ${\cal P}(\hat{H})\propto
 \exp{(-N\Tr V(\hat{H}))}$, with $V(\hat{H})$ being an even polynomial.
The corresponding eigenvalue density turns out to be highly nonuniversal
and determined by the particular form of the potential $V(H)$ \cite{BIPZ,MPS}.
Only for $V(\hat{H})=\hat{H}^2$ it is given by the
semicircular law, Eq.(\ref{semi}). Moreover, one can easily
 have a non-semicircular
eigenvalue density even for real symmetric matrices
$\hat{S};\quad S_{ij}=S_{ji}$
with i.i.d. entries, if one keeps
the mean number of non-zero entries $p$ per column to be of the order
of unity when performing the limit $N\to\infty$.
This is a characteristic feature of the
so-called {\it sparse} random matrices\cite{spa,MF,FSC}.

Much more profound universality emerges on the local scale in
the real spectra. The statistical behavior of eigenvalues
separated by distance $S=s\Delta$ measured in units
of the mean eigenvalue spacing $\Delta $ is dictated by the
global matrix symmetries (e.g. if they are complex  Hermitian or real
symmetric
\cite{Mehta}), being the same for all random matrix
ensembles within
a fixed symmetry class.
All ensemble specific information is encoded
in $\Delta $. On different levels of rigor, this
universality was established for ``invariant'' ensembles (i.e.\
matrices with invariant probabiltity distributions)
\cite{Pas1,bz,HW} and for matrices with i.i.d.\ entries, including
sparse matrices \cite{MF,note}. Similar universality
holds on a larger scale $S \gg \Delta $ \cite{univ1,kkp} and in the
vicinity of the spectrum edges \cite{bb,sosh}.

It turns out, that it is the {\it local scale} universality
 that is mostly relevant for real physical systems\cite{Bohigas}.
Namely, statistics of highly excited
bound states of {\it closed} quantum chaotic systems
of quite different microscopic nature turn out to be independent of
the microscopic details when sampled on the energy intervals large in
comparison with the mean level separation, but smaller
than the energy scale related by the Heisenberg uncertainty
principle to the
relaxation time necessary for a classically chaotic system to reach
equilibrium in phase space \cite{AlSi}.
Moreover,
these statistics
turn out to identical to
those of
large random matrices on the {\it local} scale,
with different symmetry classes
corresponding to presence or absence of time-reversal symmetry.

One of the aims of the present paper is to demonstrate that
complex spectra of weakly non-Hermitian random matrices
 possess a universality property
which  is as robust as the above mentioned local scale
universality in the real spectra of Hermitian matrices.
Weakly non-Hermitian matrices
appear naturally when one uses the Heidelberg approach to describe
few-channel chaotic scattering \cite{FS}.
When the number
$M$ of open channels is small in comparison with the number
$N$ of the  relevant resonances, the majority of  the  S-matrix poles
(resonances) are situated close to the real axis.
This is well captured within the Heidelberg approach.
With a proper normalization of $\hat H$ and $\hat W$, the imaginary part
of typical eigenvalues of the effective
Hamiltonian ${\cal H}_{eff}$ is of the order of the mean
separation between neighboring eigenvalues along the real axis.
This latter property is a characteristic feature of the regime of weak non-Hermiticity.

Motivated by this example we introduced in \cite{FKS1} another
ensemble of weakly non-Hermitian random matrices. This ensemble consists of
almost-Hermitian matrices which interpolate
between the Gaussian ensemble of Hermitian matrices (GUE) and the
Gaussian ensemble of complex matrices studied by Ginibre. It turned out that
the eigenvalue distribution for almost-Hermitian random matrices
is described by a formula \cite{FKS1} containing as two opposite
limit cases both the Wigner semicircular distribution of real
eigenvalues and the uniform distribution of complex eigenvalues obtained
by Ginibre. Further studies of almost-Hermitian random
matrices \cite{FKS2} showed that actually all their
local scale eigenvalues statistics
describe crossover between those of the GUE and Ginibre ensembles.
Later on Efetov, in his studies of directed localization
\cite{Efnonh}, discovered that weakly non-Hermitian matrices
are relevant to the problem of motion of flux lines in
superconductors with columnar defects. Efetov's matrices
are real almost-symmetric. They
interpolate between Gaussian ensemble of real symmetric
matrices (GOE) and the Gaussian ensemble of real asymmetric
matrices. This development clearly shows that, apart from being
a rich and largely unexplored mathematical object, weakly
non-Hermitian random matrices enjoy direct physical applications
and deserve a detailed study.

The present paper consists of two parts. In the first
part we study a three parameter family of
random matrix ensembles which contains the above
mentioned ensembles of almost-Hermitian and
almost-symmetric matrices.
Our random matrices are of the form
\[
\hat H =
(\hat S_1 + iu\hat A_1) + iv (\hat S_2 + iw\hat A_2),
\]
where the four matrices on the right-hand side are
mutually independent, with $\hat S_{1,2}$ being real
symmetric and $\hat A_{1,2}$ being
real skew-symmetric. By choosing matrix distributions
and varying the parameter values one obtains different
ensembles of non-Hermitian matrices. We use that
normalization of matrix elements which ensures that
\[
\lim_{N\to \infty } \frac{1}{N} \langle \Tr S_j^2 \rangle =
 \lim_{N\to \infty } \frac{1}{N} \langle \Tr A_jA_j^T \rangle =1,
 \;\;\; j=1,2
\]
$N$ being the matrix dimension. The parameters $v$ and $v$
are scaled with matrix dimension:
\[
v=\frac{\alpha }{2\sqrt{N}},\;\;\;\; u=\frac{\phi }{2\sqrt{N}},
\]
and $\alpha$, $\phi $, and $w$ are
assumed to be of the order of unity
in the limit $N\to \infty $. The above scaling of $v$
provides access to
the regime of weak non-Hermiticity, while scaling
$u$ we describe the crossover between the GOE and GUE
types of behavior of eigenvalues  of the
Hermitian part of $\hat H$.
A simple argument \cite{FKS1} based on
the perturbation theory shows that
for our random matrices
the eigenvalue deviations
from the real axis
are of the order of $1/N$ when $N$ is
large, i.e. it is of the same order as typical separation between
real eigenvalues of the Hermitian $\hat{S}_1+iu \hat{A}_1$. Hence,
in order to obtain a nontrivial eigenvalue distribution in the limit $N\to \infty $
one has to magnify the imaginary part scaling it with
the matrix dimension.

Our study of the scaled eigenvalues of $\hat H$
is based on the supersymmetry technique.  We express
the density of the scaled eigenvalues in
the form of
a correlation function of a certain zero-dimensional
non-linear $\sigma$-model. The obtained correlation function
is given by a supersymmetric integral which involves only
the density of the limit eigenvalue
distribution of the Hermitian part
of $\hat H$ and the parameters $\alpha$, $\phi$, $w$.
In two particular cases this supersymmetric integral can be
explicitly evaluated yielding the earlier obtained distributions
of complex eigenvalues for almost-Hermitian \cite{FKS1,FKS2}
and almost-symmetric matrices \cite{Efnonh}.

The supersymmetric $\sigma$-model
was invented long ago by Efetov in the context of theory of disordered
metals and the Anderson localization
and since then have been successfully applied to
diverse problems\cite{Efbook,my}.
Application of this
technique to the calculation of the mean density of complex
eigenvalues of non-Hermitian random matrices
was done for the first time in our earlier works \cite{FS,FSR,FKS1}
and further advanced by Efetov \cite{Efnonh} in the context of
 description of flux line motion in a disordered
superconductor with columnar defects.

A detailed account of our calculations is given
for sparse matrices \cite{spa,MF,FSC} with i.i.d.\ entries,
although our results are
extended to ``invariant'' ensembles
and conventional
random matrices with i.i.d.\ entries.
We assume that
matrix entries of
$\hat S_k$ and $\hat A_k$ are distributed on the
real axis with the density
\begin{equation}\label{distr}
{\cal P}(x)=\left(1-\frac{p}{N}\right)\delta (x)+\frac{p}{N}h(x),
\end{equation}
where $h(x)$ is arbitrary
symmetric density function,
$h(x)=h(-x)$, having no delta function singularity
at $x=0$ and satisfying the condition $\int x^2 h(x) dx< \infty $.
We also assume that the mean number of nonzero entries $p$
exceeds some threshold value: $p>p_l$, see \cite{note}.

We want to stress that Eq. (\ref{distr}) describes the most general class ofrandom matrices whose entries are i.i.d. variables with finite second
moment\cite{MF}. In particular, in the first part of our paper we do not assume
the matrix entries to be Gaussian.

We believe that here the power of the supersymmetry method is the most evident
and we are not aware of any other analytical technique allowing to treat
this general case non-perturbatively.

Although giving an important insight into the problem, the supersymmetry
non-linear $\sigma-$model technique
 suffers from at least two deficiencies. The most essential one is
that the
present state
of art in the application of the supersymmetry technique gives
little hope
of access to quantities describing {\it correlations}
between  different eigenvalues in the complex plane
 due to insurmountable technical difficulties.
At the same time, conventional theory of random Hermitian matrices
suggests
that  these {\it universal} correlations are the most interesting
features.
The second drawback is conceptual: the supersymmetry technique
itself is not
a rigorous mathematical tool at the moment and should be considered
as a
heuristic one from the point of view of a mathematician.

In the second part of the present paper we
develop the rigorous mathematical theory of weakly non-Hermitian
random matrices of a particular type:
 almost-Hermitian Gaussian. Our consideration is based on the method of
orthogonal polynomials. Such a method is free from the above mentioned
problem and allows us to study
correlation properties of complex spectra to the same  degree as
 is typical for earlier studied classes of random matrices.
The results were reported earlier in a form of
Letter-style communication\cite{FKS2}. Unfortunately, the paper
\cite{FKS2} contains a number of misleading misprints. For this
reason we
indicate those misprints in the present text by using footnotes.

\section{Regime of Weak non-Hermiticity:
Universal Density of Complex Eigenvalues}

To begin with, any $N\times N$ matrix $\hat{J}$ can be decomposed
into a sum of its Hermitian and skew-Hermitian parts:
$
\hat{J}=\hat{H}_1+i\hat{H}_2,
$
where  $\hat{H}_1=(\hat{J}+\hat{J}^\dagger )/2 $ and $\hat{H}_2
=(\hat{J}-\hat{J}^\dagger )/2i$. Following this, we
consider an ensemble of random $N\times N$ complex matrices
$\hat{J}=\hat{H}_1+iv\hat{H}_2$
where $\hat{H}_p;\,\, p=1,2$ are both Hermitian:
$\hat{H}^{\dagger}_p=\hat{H}_p$. The parameter $v$ is used to
control the
degree of non-Hermiticity.

In turn, complex Hermitian matrices $\hat{H}_p$
can always be represented as $\hat{H}_1=\hat{S}_1+iu\hat{A}_1$ and
$\hat{H}_2=\hat{S}_2+iw\hat{A}_2$, where $\hat{S}_p=\hat{S}_p^T$
is a real symmetric matrix, and $\hat{A}_p=-\hat{A}_p^T$
is a real antisymmetric one. From this point of view the parameters
$u,w$ control the degree of being non-symmetric.

Throughout the paper we consider the matrices
$\hat{S}_1,\hat{S}_2,\hat{A}_1,
\hat{A}_2$ to be mutually statistically independent, with i.i.d. entries
normalized in such a way that:
\begin{equation}\label{norm}
\lim_{N\to\infty}\frac{1}{N}{\mbox
Tr}\hat{S}_p^2=\lim_{N\to\infty}\frac{1}{N}{\mbox
Tr}\hat{A}_p\hat{A}_p^T=1
\end{equation}

As is well-known\cite{Bohigas}, this normalisation ensures,
that for any value of the parameter $u\ne 0$ , such that $u=O(1)$
when $N\to \infty$, statistics
of real eigenvalues of the Hermitian matrix of the form
$\hat{H}=\hat{S}+iu\hat{A}$ is identical (up to a trivial rescaling) to
that of $u=1$, the latter case known as the Gaussian Unitary
Ensemble (GUE).
On the other hand, for $u\equiv 0$ real eigenvalues of real
symmetric matrix
$S$
follow another pattern of the so-called Gaussian Orthogonal
Ensemble (GOE).

The non-trivial crossover between GUE and GOE types of statistical
behaviour happens on a scale $u\propto 1/N^{1/2}$\cite{cross}.
This scaling can be easily
understood by purely perturbative arguments\cite{Alt}.
Namely, for $u\propto 1/N^{1/2}$ the typical shift
$\delta \lambda$ of eigenvalues of the symmetric matrix $S$ due to
antisymmetric perturbation  $iu\hat{A}$ is of the same
order as the mean spacing $\Delta $ between unperturbed eigenvalues
: $\delta\lambda\sim \Delta\sim 1/N$.

Similar perturbative arguments show\cite{FKS1}, that the
most interesting behaviour
of {\it complex} eigenvalues of non-Hermitian matrices
should be expected for the parameter $v$ being scaled in
a similar way: $v\propto 1/N^{1/2}$.
It is just the regime when the {\it imaginary} parts
$\im  Z_k $ of a typical eigenvalue $Z_k$ due to
non-Hermitian perturbation is of the same order as the
mean spacing $\Delta $ between unperturbed real eigenvalues :
$\im Z_k \sim \Delta\sim 1/N$.
Under these conditions a non-Hermitian matrix $J$ still
"remembers" the statistics of its Hermitian part $\hat{H}_1$.
As will be clear afterwards, the parameter $w$
should be kept of the order of unity in order to influence the statistics
of the complex eigenvalues.

It is just the regime of {\it weak non-Hermiticity} which we are
interested in.
Correspondingly, we scale the parameters as
\footnote{In the Letter \cite{FKS2} there is a misprint in the
definition of the parameter $\alpha$.}:
\begin{equation}\label{scale}
v=\frac{\alpha}{2\sqrt{N}};\quad u=\frac{\phi}{2\sqrt{N}}
\end{equation}
and consider $\alpha,\phi,w$ fixed of the order O(1) when $N\to \infty$.

One can recover the spectral density
\begin{equation}\label{defden}
\rho(Z)=\sum_{k=1}^N\delta^{(2)}(Z-Z_k)=\sum_{k=1}^N
\delta(X-X_k)\delta(Y-Y_k)=\rho(X,Y)
\end{equation}
of complex eigenvalues $Z_k=X_k+iY_k, \quad k=1,2,...,N$
from the generating function (cf.\cite{neural,FSR})
\begin{equation}\label{genf}
{\Large \cal Z}=\frac{{\mbox
Det}\left[(Z-J)(Z-J)^{\dagger}+\kappa^2\right]}
{{\mbox Det}\left[(Z_b-J)(Z_b-J)^{\dagger}+\kappa^2\right]}
\end{equation}
as

\[
\rho(Z)=-\frac{1}{\pi}\lim_{\kappa\to 0}\frac{\partial}{\partial Z^*}
\lim_{Z_b\to Z}\frac{\partial}{\partial Z_b}{\Large\cal Z}.
\]

To facilitate the ensemble averaging we first represent the ratio
of the two
determinants in Eq.(\ref{genf}) as the Gaussian integral
\begin{equation}\label{genf1}
{\Large \cal Z}=\int \prod_{i=1}^{N}[d \Phi_{i}] \exp\{{\cal
L}_1(\Phi)+{\cal L}_2(\Phi)\}
\end{equation}
over 8-component supervectors $\Phi_{i}$,
\[
\Phi_{i}=
\left(
\begin{array}{c}
\Psi_{i}(+)\\
\Psi_{i}(-)
\end{array}
\right)
,
\Psi_{i}(\pm)=
\left(
\begin{array}{c}
\vec{R}_{i}(\pm)\\
\vec{\eta}_{i}(\pm)
\end{array}
\right),
\vec{R}_{i}(\pm)=
\left(
\begin{array}{c}
r_{i}(\pm)\\
r_{i}^{*}(\pm)
\end{array}
\right),
\vec{\eta}_{i}(\pm)=
\left(
\begin{array}{c}
\chi_{i}(\pm)\\
\chi_{i}^{*}(\pm)
\end{array}
\right)
\]
with components $r_{i}(+),r_{i}(-);\quad i=1,2,...,N$
being complex commuting
variables and $\chi_{i}(+),\chi_{i}(-)$ forming the
corresponding Grassmannian
parts of the supervectors $\Psi_{i}(\pm)$.
The terms in the exponent of Eq.(\ref{genf1}) are of the following form:
\begin{eqnarray}\label{l2}
{\cal L}_1(\Phi)&=&-\frac{i}{2}\sum_{i} \Phi_{i}^{\dagger}
\left\{S_{1,{ii}}\hat{\Lambda}
\right\}\Phi_{i}-i\sum_{i<j}
\Phi_{i}^{\dagger}\left\{S_{1,ij}\hat{\Lambda}\right\}\Phi_{j} \\
\label{l1}
{\cal L}_2(\Phi)&=&\frac{i}{2}\sum_{i} \Phi_{i}^{\dagger}
\left\{ X_b\hat{\Lambda}_b +X\hat{\Lambda}_f
-i\kappa \hat{I}+iY_b\hat{\Sigma}_{\tau,b}+iY
\hat{\Sigma}_{\tau,f}\right\}\Phi_{i}-\\
\nonumber
 & & i\sum_{i<j}\Phi_{i}^{\dagger}\left\{ivS_{2,ij}
\Sigma_{\tau}+iu A_{1,ij}
\hat{\Lambda}_{\tau}+vw A_{1,ij}\hat{\Sigma}\right\}\Phi_{j}.
\end{eqnarray}
Here
$\hat{I}_2=\diag (1,1)$,
$\hat{I_4}= \diag (\hat{I}_2,\hat{I}_2)$,
$\hat{\Lambda}= \diag (\hat{I}_4,-\hat{I}_4)$,
$\hat{\Lambda}_b= \diag (\hat{I}_2,\hat{0}_2,-\hat{I}_2,\hat{0}_2)$,
$\hat{\Lambda}_f=\hat{\Lambda}-\hat{\Lambda}_b$,
$\hat{\Sigma}=\hat{\Sigma}_b+\hat{\Sigma}_f$,
and
\[
\hat{\Sigma}_b=
\left(
\begin{array}{cc}
\hat{0}_4 &
\left(
   \begin{array}{cc}
\hat{I}_2&\hat{0}_2\\
\hat{0}_2&\hat{0}_2
   \end{array}\right)
\\
\left(
   \begin{array}{cc}
-\hat{I}_2&\hat{0}_2\\
\hat{0}_2&\hat{0}_2
   \end{array}
\right)&{\hat{0}_4}
\end{array}
\right); \quad \quad
\hat{\Sigma}_f=
\left(
\begin{array}{cc}
\hat{0}_4 & \left(
   \begin{array}{cc}
\hat{0}_2&\hat{0}_2\\
\hat{0}_2&\hat{I}_2
   \end{array}\right)\\
\left(
   \begin{array}{cc}
\hat{0}_2&\hat{0}_2\\
0&-\hat{I}_2
   \end{array}
\right)
&  \hat{0}_4
\end{array}
\right).
\]
and the matrices $\hat{\Sigma}_{\tau,b},\hat{\Sigma}_{\tau,f},
\hat{\Sigma}_{\tau},\hat{\Lambda}_{\tau}$ are obtained from the
corresponding matrices without subindex $\tau$ by replacing
all $\hat{I}_2$ blocks with the matrices
$ \hat{\tau}=\diag (1,-1)$.

We also use $\vec{\eta}_{i}^{\dagger}(\pm)=
\left(\chi_{i}^{*}(\pm);- \chi_{i}(\pm)\right)$.
When writing ${\cal L}_2(\Phi)$ in Eq.(\ref{l1})
we have used the fact that
diagonal matrix elements $S_{2,ii}$
for $ i=1,..,N$ give total contribution
of the order of $O(1/N)$ with respect to
the total contribution of the
off-diagonal ones and can be safely disregarded.

Now we should perform the ensemble averaging of the generating function.
We find it to be convenient to average first over the distribution
of matrix
elements of the real symmetric matrix $\hat{S}_1$.

These elements are
assumed to be  distributed according to Eq.(\ref{distr}).
Before presenting the derivation for our case, let us remind the general strategy. The procedure consists of three steps. First step is the
averaging of the generation function over the disorder. It can be done trivially due to statistical independence of the matrix elements in view of the integrand being a product of exponents, each  depending only
on the particular matrix element $H_{ij}$.  This averaging performed, the integrand ceases to be the simple Gaussian and thus the integration over
the supervectors can not be performed any longer without further tricks.When matrix elements are Gaussian-distributed, this difficulty is circumvented
in a standard way by exploiting the so-called Hubbard-Stratonovich transformation. That transformation amounts to making the integrand to be Gaussian with respect to components of the supervector by introducing new auxilliary integrations.
After that the integral over supervectors can be performed exactly, and remaining auxilliary degrees of freedom are integrated out in the saddle-point approximation justified by large parameter $N$.

As is shown in the paper \cite{MF}, there exists an analogue of the Hubbard-Stratonovich transformation allowing to perform the steps above
also for the case of arbitrary non-Gaussian distribution. The main difference with the Gaussian case is that the auxilliary integration has to be chosen
in a form of  a {\it functional integral }.

Our presentation follow  the procedure suggested in \cite{MF},
and presented also in some detail in \cite{FSC}

\footnote{Similarly to the paper \cite{MF} we first disregard
necessity for the compactification and use the matrix $\hat{\Lambda}$
rather than two different
matrices $\hat{L}$ and $\hat{\Lambda}$, see discussion in \cite{FSC}}.

 Exploiting the large
parameter $N\gg 1$ one can write:
\begin{equation}\label{ave}
\left\langle \exp{{\cal L}_1(\Phi)}\right\rangle\mid_{N\gg 1}\approx
\exp{\left[\frac{p}{2N}\sum_{i,j} h_{F}
(\Phi_{i}^{\dagger}\hat{\Lambda}\Phi_{j})\right]}
;\quad h_{F}(z)=\int\limits_{-\infty}^{\infty}ds h(s)e^{-isz}-1
\end{equation}

In order to proceed further we employ the functional Hubbard-Stratonovich
transformation introduced in \cite{MF}:
\begin{eqnarray}\label{HS}
\lefteqn{
\exp{
\left[
\frac{p}{2N}\sum_{i,j} h_{F}(\Phi_{i}^{\dagger}
\hat{\Lambda}\Phi_{j})
\right]
    }
=} \\
\nonumber
 & & \int {\cal D}g \
\exp{
\left[
-\frac{pN}{2}\int d\Theta d\tilde{\Theta} g(\Theta)
C(\Theta,\tilde{\Theta})g(\tilde{\Theta})+p\sum_{i=1}^{N}
g(\Phi_{i})
\right]
    }
\end{eqnarray}
where the kernel $C(\Theta,\tilde{\Theta})$ is determined by the
relation:
\begin{equation}\label{ker}
\int d\tilde{\Theta}C(\Theta,\tilde{\Theta})
h_{F}(\tilde{\Theta}\hat{\Lambda}\Phi)=
\delta(\Theta,\Phi)\end{equation}
with the right-hand side of the eq.(\ref{ker}) being the
$\delta-$function
in the space of supervectors.

Substituting eq.(\ref{HS}) into averaged eq.(\ref{genf1}) and
changing the order of
integrations over $[d\Phi]_{i}$ and ${\cal D} g$ one obtains the averaged
generating function in the form:
\begin{equation}\label{gav}
\left\langle{\cal Z}\right\rangle=\int {\cal D}g\exp{\left[-N{\cal L}(g)+
\delta {\cal L}(g)\right]}
\end{equation}
where
\begin{eqnarray}\nonumber
{\cal L}(g)&=&\frac{p}{2}\int [d\Phi][d\tilde{\Phi}]\
g(\Phi)C(\tilde{\Phi},
\Phi)g(\Phi)-\ln{\int [d\Phi]\  e^{{\cal F}(\Phi)}}\\
\label{calg}
\delta {\cal L}(g)&=&\ln \
\frac{\int\prod_{i=1}^{N}[d\Phi_{i}]\
\exp{\left[
\sum_{i}{\cal F}(\Phi_{i})+R\{\Phi\}\right]}}{\int
\prod_{i=1}^{N}[d\Phi_{i}]\
\exp{\left[\sum_{i}{\cal F}(\Phi_{i})\right]}}
\end{eqnarray}
with ${\cal F}(\Phi)
=\frac{i}{2}X\Phi^{\dagger}\hat{\Lambda}\Phi+pg(\Phi)$,
\[
R\{\Phi\}=\frac{1}{2}\sum_{i=1}^N\Phi_i^{\dagger}\hat{f}\Phi_i
-i\sum_{i<j}\Phi_{i}^{\dagger}\left\{ivS_{2,ij}\hat{\Sigma}_{\tau}+iu
A_{1,ij}
\hat{\Lambda}_{\tau}+
vw A_{1,ij}\hat{\Sigma}\right\}\Phi_{j},
\]
and $\hat{f}=\kappa\hat{I}-i(X-X_b)\hat{\Lambda}_b-
Y_b\hat{\Sigma}_{\tau,b} -Y\hat{\Sigma}_{\tau,f}$.

We are interested in evaluating the functional integral over ${\cal D}g$
in the limit $N\to \infty$ and  $\kappa\to 0; X\to X_b$. Moreover, we
expect eigenvalues of weakly non-Hermitian
matrices to have imaginary parts $Y$
to be of the order of $1/N$. Remembering also the chosen scaling
(\ref{scale}), we conclude that the argument of the logarithm in
Eq.(\ref{calg}) is close to unity and the term $\delta{\cal
L}(\Phi)$ in
Eq.(\ref{gav})
should be treated
as a small perturbation to the first one. Then the functional integral
of the type $\int {\cal D}g (...) \exp{-N{\cal L}(g)}$ can be
evaluated by the saddle-point method. Variating the "action"
${\cal L}(g)$ and using  the relation
eq.(\ref{ker}) one obtains the following saddle point equation
$\delta {\cal
L}(g)/\delta g=0$ for the function
$g(\Phi)$:
\begin{equation}\label{sp}
g(\Phi)=\frac{\int [d\Theta] h_{F}(\Phi^{\dagger}\hat{L}\Theta)
\exp{{\cal F}(\Theta)}}{\int [d\Theta]\exp{{\cal F}(\Theta)}}
\end{equation}

A quite detailed investigation of the
properties of this equation was performed
in \cite{MF,FSC}. Below we give a summary of the main features of the
eq.(\ref{sp}) following from such an analysis.

First, the solution $g(\Phi)$ to this equation can be sought for in
a form
of a function $g_{s}(\Phi)=g_{0}(x,y)$ of two superinvariants:
$x=\Phi^{\dagger}\Phi$ and $y=\Phi^{\dagger}\hat{\Lambda}\Phi$.

As the result, the denominator in eq.(\ref{sp}) is equal to $1$ due to
the identity $\int [d\Phi] F(x,y)=F(0,0)$ which is a particular case of
the so-called Parisi-Sourlas-Efetov-Wegner (PSEW) theorem, see
e.g.\cite{my} and
references therein.
However, the form of the function $g_{0}(x,y)$ is essentially different
for the number of nonzero elements $p$ per matrix column
exceeding the threshold value $p=p_{l}$  and for $p<p_{l}$
\cite{MF,note}. Namely, for $p>p_{l}$ the function $g_{0}(x,y)$
is an {\it analytic} function of {\it both} arguments $x$ and $y$,
whereas
for $p<p_{l}$ such a function is dependent only on the second argument
$y=\Phi^{\dagger}\hat{\Lambda}\Phi$. At the same time, the
saddle-point equation
eq.(\ref{sp}) is {\it always} invariant w.r.t. any transformation
$g(\Phi)\to g(\tilde{T}\Phi)$ with supermatrices $\tilde{T}$ satisfying
the condition $\tilde{T}^{\dagger}\hat{\Lambda}\tilde{T}=\hat{\Lambda}$.

Combining all these facts together one finds,
that for $p>p_{l}$ a saddle-point
solution $g_{s}(\Phi)$ gives rise to the whole continuous manifold of
saddle-point solutions of the form: $g_{T}(\Phi)\equiv
g_{s}(\tilde{T}\Phi)=
g_{0}(\Phi^{\dagger}\tilde{T}\tilde{T}\Phi,\Phi^{\dagger}\hat{\Lambda}
\Phi)$, so that all the manifold gives a nonvanishing contribution to the
functional integral eq.(\ref{gav}). It is the existence of the
saddle-point
manifold that is actually responsible for the universal random-matrix
correlations\cite{MF}.

We see, that the saddle-point manifold is parametrized by the
supermatrices
$\tilde{T}$. It turns out, however, that one has to "compactify"
the manifold
of $\tilde{T}$ matrices with respect to the "fermion-fermion"
block in order to ensure
convergence of the integrals over the saddle-point manifolds\cite{VWZ}.
The resulting "compactified" matrices $\hat{T}$ form a graded Lie group
$\mbox{UOSP}(2,2/4)$. Properties of such matrices can
be found in \cite{VWZ} together with the integration measure $d\mu(T)$.

>From now on we are going to consider only the case $p>p_{l}$. The
program of
calculation is as follows: (i) To find the expression for the term
$\delta {\cal L}(g)$ on the saddle-point manifold $g=g_{T}(\Phi)$
 in the limit $N\to \infty$ and (ii) to calculate the integral
 over the saddle-point manifold exactly.

Expanding the expression in Eq.\ (\ref{calg}) to the
first non-vanishing
order in $\hat{f},\hat{S}_2,\hat{A}_1,\hat{A}_{2}$, introducing the
notation ${\cal F}_T(\Phi)=
\frac{i}{2}X\Phi^{\dagger}\hat{\Lambda}\Phi+pg_{T}(\Phi)$
and using the relations
\begin{equation}\begin{array}{c}
\int [d\Phi] \exp{{\cal F}_{T}(\Phi)}=1;
\quad \int \prod_{k=1}^{N}[d\Phi_{k}]
\left(\Phi^{\dagger}_{i}\hat{B}\Phi_{j}\right)|_{i<j}
\exp{\sum_{k=1}^{N}{\cal F}_{T}(\Phi_{k})}=0;\\ \\
 \int \prod_{k=1}^{N}[d\Phi_{k}]
\left(\Phi^{\dagger}_{i_{1}}\hat{B}\Phi_{j_{1}}\right)|_{i_{1}<j_{1}}
\left(\Phi_{i_{2}}^{\dagger}\hat{C}\Phi_{j_{2}}\right)|_{i_{2}<j_{2}}
\exp{\sum_{k=1}^{N}{\cal F}_{T}(\Phi_{k})}\\=
\delta_{i_{1}i_{2}}\delta_{j_{1}j_{2}}
\int [d\Phi_{i_{1}}][d\Phi_{j_{1}}]
\left(\Phi^{\dagger}_{i_{1}}\hat{B}\Phi_{j_{1}}\right)
\left(\Phi_{i_{1}}^{\dagger}\hat{C}\Phi_{j_{1}}\right)_{i_{1}<j_{1}}
\exp{\left[{\cal F}_{T}(\Phi_{i_{1}})+{\cal F}_{T}(\Phi_{j_{1}})\right]}
\end{array}\end{equation}
which hold for arbitrary $8\times 8$ supermatrices $\hat{B},\hat{C}$,
one finds that
\begin{equation}\label{expan}
 \delta {\cal L}(g_{T}) =\frac{1}{2}\langle\Phi^{\dagger}
\hat{f}\Phi\rangle_T+\delta{\cal L}_R+\delta{\cal L}_{IR},
\end{equation}
where
\begin{eqnarray*}
\delta{\cal L}_R &=& \frac{v^2}{2}
\sum_{i<j}\left(\hat{S}_2\right)^2_{ij}
\left\langle\left[\Phi_1^{\dagger}
\hat{\Sigma}_{\tau}\Phi_2\right]
\left[\Phi_1^{\dagger}\hat{\Sigma}_{\tau}
\Phi_2\right]\right\rangle_T  \\
 &+& \frac{u^2}{2}
\sum_{i<j} \left(\hat{A}_1\right)^2_{ij}
\left\langle\left[\Phi_1^{\dagger}
\Lambda_{\tau}
\Phi_2\right]\left[\Phi_1^{\dagger}\hat{\Lambda}_{\tau}
\Phi_2\right]\right\rangle_T - \frac{v^2w^2}{2}
\sum_{i<j} \left(\hat{A}_2\right)^2_{ij}
\left\langle\left[\Phi_1^{\dagger}
\hat{\Sigma}\Phi_2\right]\left[\Phi_1^{\dagger}\hat{\Sigma}
\Phi_2\right]\right\rangle_T,\\
\delta{\cal L}_{IR} &=& uv
\sum_{i<j} \left(\hat{S}_2\right)_{ij}\left(\hat{A}_1\right)_{ij}
 \left\langle\left[\Phi_1^{\dagger}
\hat{\Sigma}_{\tau}\Phi_2\right]
\left[\Phi_1^{\dagger}\hat{\Lambda}_{\tau}
\Phi_2\right]\right\rangle_T - iv^2w
\sum_{i<j}\left(\hat{S}_2\right)_{ij}\left(\hat{A}_2\right)_{ij}
\left\langle\left[\Phi_1^{\dagger}\hat{\Sigma}_{\tau}
\Phi_2\right]\left[\Phi_1^{\dagger}\hat{\Sigma}
\Phi_2\right]\right\rangle_T  \\
 &-& iuvw
\sum_{i<j}\left(\hat{A}_1\right)_{ij}\left(\hat{A}_2\right)_{ij}
\left\langle\left[\Phi_1^{\dagger}
\hat{\Lambda}_{\tau}\Phi_2\right]\left[\Phi_1^{\dagger}
\hat{\Sigma}\Phi_2\right]\right\rangle_T
\end{eqnarray*}
and we used the notations:
\begin{eqnarray*}
\left\langle\left[\Phi^{\dagger}\hat{B}\Phi\right]\right\rangle_T
&=&
\int [d\Phi]\left(\Phi^{\dagger}\hat{B}\Phi\right)e^{{\cal F}_T(\Phi)} \\
\left\langle\left[\Phi_1^{\dagger}\hat{B}
\Phi_2\right]\left[\Phi_1^{\dagger}\hat{C}\Phi_2\right]\right\rangle_T
&=&
\int [d\Phi_{1}][d\Phi_2]
\left(\Phi^{\dagger}_{1}\hat{B}\Phi_{2}\right)
\left(\Phi_1^{\dagger}\hat{C}\Phi_{2}\right)
\exp{\left[{\cal F}_{T}(\Phi_{1})+{\cal F}_{T}(\Phi_{2})\right]}.
\end{eqnarray*}

It is clear that with the chosen normalization [see
Eqs.\ (\ref{norm}) -- (\ref{scale})] we have
\begin{equation}
v^2\sum_{i<j}
\left(\hat{S}_2\right)^2_{ij}\to\frac{\alpha^2}{8};\quad
u^2\sum_{i<j}
\left(\hat{A}_1\right)^2_{ij}\to\frac{\phi^2}{8};\quad
w^2v^2\sum_{i<j}
\left(\hat{A}_2\right)^2_{ij}\to\frac{w^2\phi^2}{8}
\end{equation}
when $N\to \infty$. On the other hand, it is easy to see that:
\begin{equation}
uv\sum_{i<j}
\left(\hat{S}_2\right)_{ij}\left(\hat{A}_1\right)_{ij}
\sim wv^2\sum_{i<j}\left(\hat{S}_2\right)_{ij}
\left(\hat{A}_2\right)_{ij}\sim
wuv\sum_{i<j}\left(\hat{A}_1\right)_{ij}
\left(\hat{A}_2\right)_{ij}=O(\frac{1}{N})
\end{equation}
because of the statistical independence of
$\hat{S}_2,\hat{A}_1,\hat{A}_2$ and the chosen normalization of
matrix elements.
Therefore, the part  $\delta{\cal L}_{IR}$ can be safely neglected
in the limit of large $N$.

To proceed further it is convenient to introduce the $8\times 8$
supermatrix $W$ with elements
\begin{equation}\label{w}
W_{\alpha\beta}\equiv \int [d\Phi]\
\Phi_{\alpha}\Phi_{\beta}^{\dagger}
e^{{\cal F}_{T}(\Phi)}\end{equation}
Exploiting the saddle-point equation for the function $g_T(\Phi)=
g_0\left(\Phi^{\dagger}\hat{T}^{\dagger}\hat{T}\Phi,\Phi^{\dagger}
\Lambda\Phi\right)$ one can show (details can be found in
\cite{FSC}, Eqs.(67-70)) that the supermatrix $\hat{W}$ whose
elements are
$W_{\alpha\beta}$  can be written as:
\begin{equation}\label{ww}
\hat{W}=\frac{2}{B}
\left[g_{0y}\hat{\Lambda}+ig_{0x}\hat{\Lambda}\hat{Q}
\right]
\end{equation}
where $B$ is the second moment of the distribution $h(s)$:
$B=\int h(s) s^2 ds$
and $g_{0x}=\partial g_{0}/\partial x|_{x=y=0};\quad g_{0y}=
\partial g_{0}/\partial y|_{x=y=0}$.
In the expression above we introduced a new supermatrix:
$\hat{Q}=-i\hat{T}
^{-1}\hat{\Lambda}\hat{T}$.

Using the definition of the matrix $W$, one can rewrite the part
$\delta{\cal L}_R$ as follows (cf.\cite{FSC}, eqs. (71)-(73)):
\begin{eqnarray}
\delta{\cal
L}_R=-\frac{1}{16}\left[\alpha^2  \Str \hat{W}\hat{\Sigma}_{\tau}
\hat{W}\hat{\Sigma}_{\tau}
-\phi^2  \Str \hat{W}\hat{\Lambda}_{\tau}\hat{W}\hat{\Lambda}_\tau+
w^2\phi^2  \Str \hat{W}\hat{\Sigma}\hat{W}\hat{\Sigma}\right]
\end{eqnarray}
Now one can use Eq.(\ref{w}) together with the properties:
$  \Str \hat{Q}=  \Str \hat{\Lambda}=  \Str \hat{I}=0;\quad
Q^2=-\hat{I}$ to show that:
\[
\delta{\cal  L}_R
=\frac{g_{0x}^2}{4B^2}\left[\alpha^2  \Str \hat{Q}\hat{\sigma}_{\tau}
\hat{Q}\hat{\sigma}_{\tau}-\phi^2
  \Str \hat{Q}\hat{\tau}_2\hat{Q}\hat{\tau}_2+w^2\phi^2
  \Str \hat{Q}\hat{\sigma}\hat{Q}\hat{\sigma}\right]
\]
where the $8\times 8$
supermatrices entering these expressions are as follows:
\[
\hat{\tau}_2= \diag \{\hat{\tau}_3,\hat{\tau}_3\};\quad
\hat{\sigma}_{\tau}=\left(\begin{array}{cc}0&\hat{\tau}_3\\
\hat{\tau}_3&0\end{array}\right);\quad
\hat{\sigma}=
\left(
\begin{array}{cc}
0&\hat{I}_4\\
\hat{I}_4&0
\end{array}
\right)
\]
and $\hat{\tau}_3$ is $4\times 4$ diagonal,
$\hat{\tau}_3= \diag \{\hat{\tau},\hat{\tau}\}$.

In the same way one finds:
\begin{eqnarray*}
\left\langle\Phi^{\dagger}\hat{f}\Phi\right\rangle_T
&=&
-\frac{4ig_{0y}}{B}(X-X_b)+\\
& &
\frac{2ig_{0x}}{B}
\left[
\kappa  \Str \hat{Q}\hat{\Lambda}-i(X-X_b)  \Str
\hat{K}_B\hat{Q}
-Y_B   \Str \hat{\sigma}_\tau^{(B)}\hat{Q}-Y
  \Str \hat{\sigma}_\tau^{(F)}\hat{Q}
\right],
\end{eqnarray*}
where
\[
\hat{K}_B= \diag \{\hat{I}_2,\hat{0}_2,\hat{I}_2,\hat{0}_2\};\quad
\hat{\sigma}_{\tau}^{(B,F)}=
\left(
\begin{array}{cc}
\hat{0}_4&\hat{\tau}_3^{(B,F)}\\
\hat{\tau}_3^{(B,F)}&\hat{0}_4
\end{array}\right)
\]
and $\hat{\tau}_3^{B,F}$ are $4\times 4$ diagonal supermatrices:
$\hat{\tau}_3^{(B)}= \diag \{\hat{\tau},\hat{0}_2\}$ and
$\hat{\tau}_3^{(F)}= \diag \{\hat{0}_2,\hat{\tau}\}$.

At last, we use the relation between $g_0(x,y)$ and the mean eigenvalue
density  for a sparse
symmetric matrix $\hat{S}_1$ at the point $X$ on the
real axis derived in \cite{MF}:
\begin{equation}\label{meanden}
\nu(X)\equiv\frac{1}{N}\langle \rho(X)\rangle=-\frac{2}{\pi B}g_{0x}
\end{equation}

Substituting expressions for $\delta{\cal L}_R$ and
$\left\langle\Phi^{\dagger}\hat{f}\Phi\right\rangle_T$ to the
generating
function ${\cal Z}$ represented as an integral over
the saddle-point manifold
parametrized by the supermatrices $\hat{T}$ (or,
equivalently, by the supermatrices $\hat{Q}=-i\hat{T}^{-1}
\hat{\Lambda}\hat{T}$) and
performing the proper limits we finally obtain:
\begin{equation}\label{unires}
\langle\rho(X,Y)\rangle=\frac{\pi[N\nu(X)]^2}{16}
\int d\mu(\hat{Q})  \Str \left(\hat{\sigma}_\tau^{(F)}\hat{Q}\right)
  \Str \left(\hat{\sigma}_\tau\hat{Q}\right)\exp{-S(\hat{Q})}
\end{equation}

\[
S(\hat{Q})=-\frac{i}{2}y
  \Str
\left(\hat{\sigma}_\tau\hat{Q}\right)-
\frac{a^2}{16}
  \Str \left(\hat{\sigma}_\tau\hat{Q}\right)^2+
\frac{b^2}{16}
  \Str
\left(\hat{\tau}_2\hat{Q}\right)^2
-\frac{c^2}{16}
  \Str
\left(\hat{\sigma}\hat{Q}\right)^2
\]

where we introduced the scaled imaginary parts $y=\pi\nu(X)NY$
and used the notations: $a^2=\left(\pi\nu(X)\alpha\right)^2,\quad
b^2=\left(\pi\nu(X)\phi\right)^2,\quad c^2=\left(\pi\nu(X)\alpha
w\right)^2$

The expression (\ref{unires}) is just the universal $\sigma-$ model
representation of the mean density of complex eigenvalues in the
regime of weak non-Hermiticity we were looking for.
The universality is clearly manifest: all the particular details
about the
ensembles entered only in the form of mean density of
real eigenvalues $\nu(X)$.  The density of complex
eigenvalues turns out to be dependent on three parameters:
$a,b$ and $c$, controlling the degree of non-Hermiticity
($a$), and symmetry properties of the
Hermitian part ($b$) and non-Hermitian part ($c$).

The following comment is appropriate here. The derivation above
was done for ensembles with i.i.d. entries. However, one can
satisfy oneself
 that the same expression would result if one start
instead from any "rotationally invariant" ensemble of
 real symmetric matrices $\hat{S}_1$.
To do so one can employ the procedure invented by Hackenbroich and
Weidenm\"{u}ller \cite{HW} allowing one to map the
correlation functions of the invariant
ensembles (plus perturbations) to that of Efetov's $\sigma-$model.

Still, in order to get an explicit expression for the
density of complex eigenvalues one has to
evaluate the integral over the set of supermatrices
$\hat{Q}$. In general, it is an elaborate
task due to complexity of that manifold.

At the present moment such an evaluation was successfully performed
for two
important cases: those of almost-Hermitian matrices and
real almost-symmetric
matrices. The first case ( which is technically the simplest one)
corresponds to $\phi\to\infty$, that is $b\to \infty$.
Under this condition only that part of the matrix
$\hat{Q}$ which commutes with
$\hat{\tau}_2$ provides a nonvanishing contribution. As the result,
$  \Str \left(\hat{\sigma}\hat{Q}\right)^2=  \Str
\left(\hat{\sigma}_{\tau}\hat{Q}\right)^2$
so that second and fourth term in Eq.(\ref{unires})
can be combined together. Evaluating the resulting integral,
and introducing the notation $\tilde{a}^2=a^2+c^2$ one finds \cite{FKS1}:
\begin{eqnarray}\label{13}
\rho_X(y)=\sqrt{\frac{2}{\pi}}\frac{1}{\tilde{a}}\exp
 \left( -\frac{2y^2}{\tilde{a}^2} \right)
\int\limits_0^1  dt \cosh (2ty) \exp{(-\tilde{a}^2t^2/2)},
\end{eqnarray}
where $\rho_X(y)$ is the density of the scaled imaginary parts $y$
for those eigenvalues, whose real parts are situated around the point
$X$ of the spectrum. It is related to the two-dimensional density
as $\rho_X(y)=\rho(X,Y)/\pi(N\nu(X))^2$.

It is easy to see, that when $\tilde{a}$ is large one can effectively
put the upper boundary of integration in Eq.(\ref{13}) to be infinity
due to the Gaussian cut-off of the integrand. This immediately
results in the uniform density $\rho_X(y)=(\tilde{a}^2)^{-1}$ inside the
interval $|y|<\tilde{a}^2/2$ and zero otherwise. Translating this
result to the
two-dimensional density of the original variables $X,Y$, we get:
\begin{equation}\label{girko}
\rho(X,Y)=
\left\{
\begin{array}{ll}
\displaystyle{
\frac{N}{4\pi v^2(1+w^2)}
              }& \mbox{if}\,\,
\displaystyle{
|Y|\le 2\pi\nu(X)v^2(1+w^2)
             }\\
0&\mbox{otherwise}\end{array}
\right.
\end{equation}

This result is a natural generalisation of the so-called "elliptic
law" known
for strongly non-Hermitian random matrices\cite{Gin,neural}.
Indeed, the curve encircling the domain of the uniform eigenvalue
density is an ellipse: $\frac{Y^2}{2v^2(1+w^2)}+\frac{X^2}{4}=1$ as
long as the
mean eigenvalue density of the Hermitian counterpart is given by the
semicircular law, Eq.(\ref{semi}) (with the parameter $J=1$). The
semicircular density is known to be
shared by ensembles with i.i.d. entries, provided the mean
number $p$ of
non-zero elements per row grows with the matrix size as $p\propto
N^{\alpha}; \,\, \alpha>0$, see \cite{MF}. In the general case of
sparse or "rotationally invariant" ensembles the function $\nu(X)$
might be quite different from the semicircular law.
Under these conditions Eq.(\ref{girko})
still provides us with the corresponding density of complex eigenvalues.

The second nontrivial case for
which the result is known explicitly is due to Efetov\cite{Efnonh}.
It is the limit
of slightly asymmetric real
matrices corresponding in the present notations to:
$\phi\to 0; w\to \infty$ in such a way that the product
$\phi w=\tilde{c}$ is kept fixed.
The density of complex eigenvalues turns out to be given by:
\begin{eqnarray}\label{Efres}
\rho_X(y)&=&\delta(y)\int\limits_0^1 dt\exp{(-\tilde{c}^2t^2/2)}  \\
\nonumber
&+2\sqrt{\frac{2} {\pi}}&\frac{|y|}{\tilde{c}}
\int\limits_1^{\infty}du \exp \left(-\frac{2y^2u^2}{\tilde{c}^2} \right)
\int\limits_0^1  dt t \sinh (2t|y|) \exp{(-\tilde{c}^2t^2/2)},
\end{eqnarray}

The first term in this expression shows that everywhere in the regime of
"weak asymmetry" $\tilde{c}<\infty$ a finite fraction of
eigenvalues remains on the real axis.

Such a behaviour is qualitatively
different from that typical for the
case of "weak non-Hermiticity" $\tilde{a}<\infty$, where eigenvalues
acquire a nonzero imaginary part with probability one.

In the limit $\tilde{c}>>1$ the portion of real eigenvalues
behaves like $\tilde{c}^{-1}$. Remembering the normalisation
of the parameter $v$, Eq.(\ref{norm}), it is easy to see that
for the case of $v=O(1)$ the number of real eigenvalues should scale
as $\sqrt{N}$.
Indeed, as
was first noticed by
Sommers et al. \cite{neural,Lehm1}
the number of {\em real} eigenvalues of
strongly asymmetric
{\em real} matrices is proportional to
$\sqrt{N}$ . This  and the fact that the
mean density of real eigenvalues is constant
was later proved by Edelman et al. \cite{Edel}.

\section{Gaussian almost-Hermitian matrices: from Wigner-Dyson to
Ginibre eigenvalue statistics}
In the previous section we obtained the eigenvalue distribution
in the regime of weak non-Hermiticity for the random matrices
of the form $\hat J =\hat H_1 +iv \hat H_2$, with
$\hat H_1 $ and $\hat H_2$ being mutually independent Hermitian
random matrices with i.i.d.\ matrix entries. The obtained eigenvalue
distribution appeared to be universal, i.e.\ independent of the
probability
distribution of the Hermitian matrices $\hat H_1 $ and $\hat H_2$.

In the present section we reexamine a particular case of $\hat J$ when
both $\hat H_1 $ and $\hat H_2$ are taken to be Gaussian. In this
special case
not only the mean eigenvalue density but also the eigenvalue
correlation functions can be obtained and studied in great detail.

The ensemble of random matrices that will  be considered in this section
is specified by the probability measure
$d\mu (\hat J)={\cal P}(\hat J) d \hat J$,
\begin{eqnarray}\label{P(J)}
{\cal P}(\hat{J})
=\left( \frac{N}{\pi \sqrt{1-\tau^2} }\right)^{N^2}
\exp \left[
-\frac{N}{(1-\tau^2)} \Tr
\left(
\hat{J}\hat{J}^\dagger - \tau \re \hat{J}^2
\right)
\right]
\end{eqnarray}
on the set ${\cal M}$ of complex $N\times N$ matrices
with the matrix volume element
\[
d\hat J = \prod_{j,k=1}^{N}d^2 J_{jk}, \;\;
d^2 J_{jk}\equiv d \re J_{jk} d\im J_{jk}.
\]
If the Hermitian $\hat H_1$ and $\hat H_2$
are taken independently from the GUE, the probability
distribution of $\hat J =\hat H_1 +iv \hat H_2$
is described by the above-given measure $d\mu (\hat{J})$
with
\[
\tau =\frac{1-v^2}{1+v^2}
\]
provided that $\hat H_1$ and $\hat H_2$ are normalized to
satisfy $\langle \Tr \hat H_p^2 \rangle = N(1+\tau )/2$,
$p=1,2$.

The parameter $\tau$, $0 \le \tau \le 1$,
controls the magnitude of the
correlation between
$J_{jk}$ and $J_{kj}$: $\langle J_{jk}J_{kj}
\rangle = \tau /N $, hence
the degree of non-Hermiticity.
All $J_{jk}$  have zero mean and variance
$\langle |J_{jk}|^2 \rangle =1/N$ and only
$J_{jk}$ and $J_{kj}$ are pairwise correlated.
If $\tau =0$ all $J_{jk}$
are mutually independent and
we have maximum non-Hermiticity.
When $\tau $ approaches unity, $J_{jk}$ and
$J_{kj}^*$ are related via $J_{jk}=J_{kj}^*$
and we are back to the ensemble of Hermitian matrices.

Our first goal is to obtain the density of the joint distribution
of eigenvalues in the random matrix ensemble
specified by Eq.\ (\ref{P(J)}). First of all, one can disregard
the matrices whose characteristic polynomial has multiple roots.
For, the set of such matrices forms a surface in ${\cal M}$,
hence has zero volume. Every matrix off this surface has $N$
distinct eigenvalues and we label them $Z_1, \ldots , Z_N$
ordering them in such a way that
\begin{equation}\label{b0}
|Z_1| \le |Z_2| \le \ldots \le |Z_N|
\end{equation}
and if $ |Z_j| = |Z_{j+1}|$ for some $j$ then $\arg Z_j < \arg Z_{j+1}$.
Given $\hat J$, one can always find a unitary matrix $\hat U$
and a triangular $\hat T$ ($T_{jk}=0$ if $j>k$) such that
\begin{equation}\label{b1}
\hat J =\hat U \hat T \hat U^{-1}
\end{equation}
and $T_{jj}=Z_j$ for every $j$ \cite{Wilkinson}.
The choice of $\hat U$ and $\hat T$
is not unique. For, multiplying $\hat U$ to the right by a
unitary diagonal matrix  $\hat \Phi $ one can also write $\hat J =\hat V
\hat S \hat V^{-1}$, where $\hat V=\hat U \hat \Phi $ is unitary,
$\hat S$ is triangular, and again $S_{jj}=Z_j$ for every $j$.
It is natural, therefore, to impose a restriction on $\hat U$
requiring, for instance, the first non-zero element in each column
of $\hat U$ to be real positive. Then the correspondence (\ref{b1})
between $\hat J$ and $(\hat U, \hat T)$ is one-to-one.

The idea of using the decomposition  (\ref{b1}) ( which is often called
the Schur decomposition)  for derivation of the joint distribution
of eigenvalues goes back to Dyson \cite{Dyson} and we simply follow
his argument. To obtain the density of the joint distribution
one integrates (\ref{P(J)}) over the set of matrices whose
eigenvalues are $Z_1, \ldots , Z_N$. To perform the integration,
one changes the variables from $\hat J$ to $(\hat U , \hat T)$
and integrates over $\hat U$ and the off-diagonal elements of $\hat T$.
The Jacobian of the transformation $\hat J  \to (\hat U , \hat T)$
depends only on the eigenvalues and is given by the squared modulus
of the Vandermonde determinant of the $\{Z_j\}$ \cite{Dyson}.
Since $\hat U$ is unitary,
\begin{equation}\label{b2}
\Tr (\hat J \hat J^{\dagger} +\tau \re \hat J^2)=
\Tr (\hat T \hat T^{\dagger} +\tau \re \hat T^2).
\end{equation}
Therefore, the integral over $\hat U$ yields
\[
\frac{\Vol [U(N)]}{(2\pi
)^N}=\frac{\pi^{N(N-1)/2}}{\prod_{n=1}^{N-1} n!},
\]
where $\Vol [U(N)]$ is the volume of the unitary group $U(N)$.
Since $\hat T$ is triangular, the integration over the
off-diagonal entries of $\hat T$ reduces, in view of
(\ref{b2}), to the Gaussian integral
\[
\prod_{1\le j<k \le N} \int d^2 T_{jk}
\exp \left[ -\frac{N}{1-\tau^2}
\left(  |T_{jk}|^2 -\tau \re T_{jk}^2 \right)
     \right]=
\left[
\frac{\pi (1-\tau^2)}{N}
\right]^{N(N-1)/2}.
\]
Collecting the constants one obtains the desired density.
Obviously, it is symmetric in the eigenvalues $\{Z_j\}$. Therefore,
the above restriction of Eq.\ (\ref{b0})
on the eigenvalues can be removed by
reducing the obtained density in $N!$ times.
Thus finally, the density of the
joint distribution of (unlabelled) eigenvalues in the random matrix
ensemble specified by Eq.\ (\ref{P(J)}) is given by
\begin{equation}\label{P(Z)}
{\cal P}_N(Z_1, \ldots, Z_N)= \frac{N^{N(N+1)/2}}{\pi^N 1!
\cdots N! (1-\tau^2)^{N/2} }\
\prod_{j=1}^N w^2(Z_j)\
\prod_{j<k}|Z_j-Z_k|^2,
\end{equation}
where
\begin{equation}\label{b3}
w^2(Z)=\exp \left\{-\frac{N}{1-\tau^2}
\left[|Z|^2 - \frac{\tau}{2}\left(Z^2+{Z^*}^2\right) \right]  \right\}.
\end{equation}

The form of the distribution
Eq.\ (\ref{P(Z)}) allows one to employ
the powerful method of orthogonal polynomials \cite{Mehta}.
Let $H_n(z)$ denotes
$n$th Hermite polynomial,
\begin{equation} \label{H}
H_n(z)=\frac{(\pm i)^n}
{\sqrt{2\pi}}\! \exp{\left(\! \frac{z^2}{2}\! \right)}
\int\limits_{-\infty}^{\infty}\!
dt\  t^n \exp{\left(-\frac{t^2}{2}\mp izt\right)}.
\end{equation}
 These Hermite polynomials
are orthogonal on the real axis with the weight function
$\exp (-x^2 /2) $ and are determined by the following
generating function
\begin{equation}\label{b4}
\exp \left(
           zt - \frac{t^2}{2}
      \right)
= \sum_{n=1}^\infty H_n(z) \frac{t^n}{n!}.
\end{equation}
It is convenient to rescale Hermite polynomials in the following way:
\begin{eqnarray}\label{p_n}
p_n(Z)=\frac{\tau^{n/2} \sqrt{N} }
{\sqrt{\pi }\sqrt{ n!}(1-\tau^2 )^{1/4}}
H_n\left( \sqrt{ \frac{N}{\tau}}Z\right).
\end{eqnarray}
The main reason for doing that rescaling is
that these new polynomials $p_n(Z)$,
$n=0,1,2, \ldots $,
are orthogonal in the {\it complex plane}
$Z=X+iY$
with the weight function $w^2(Z)$ of Eq.\ (\ref{b3}):
\begin{equation}
\int d^2 Z\  p_n(Z)p_m(Z^*)w^2(Z)  = \delta_{nm}.
\end{equation}
(Recall that $d^2 Z = dX dY$.)
We borrowed this observation which is crucial for our analysis
from the paper \cite{orth}
(see also the related paper \cite{FJ}).
A quick check of the orthogonality relations
is possible with the help of the generating function
(\ref{b4}).

With these orthogonal polynomials in hand,
the standard machinery of the method of
orthogonal polynomials\cite{Mehta} yields
the $n$-eigenvalue correlation functions
\begin{equation}
R_n(Z_1,...,Z_n)=\frac{N!}{(N-n)!}\int d^2Z_{n+1}...d^2Z_N\
{\cal P}_N\{Z\}
\label{R_n}
\end{equation}
in the form
\begin{equation}\label{b4a}
R_n(Z_1,...,Z_n)=\det \left[ K_N(Z_j,Z_k)\right]_{j,k=1}^n,
\end{equation}
where the kernel $K_N(Z_1,Z_2)$ is given by
\begin{equation}
K_N(Z_1,Z_2)=
w(Z_1)w(Z_2^*)
\sum_{n=0}^{N-1}p_n(Z_1)p_n(Z_2^*).
\label{K}
\end{equation}
In particular, define the density of eigenvalues as in
Eq.(\ref{defden}),
so that the number of eigenvalues in domain $A$ of the complex plane
is given by the integral
\begin{equation}\label{b6}
n(A)=\int\limits_A d^2 Z\ \rho(Z).
\end{equation}
Notice that the averaged density of eigenvalues
$\langle \rho (Z) \rangle$
is simply $R_1(Z)$. From
Eqs.\ (\ref{b4a}), (\ref{K}), and (\ref{p_n}) one infers that
\begin{equation}
\label{den}
R_1(Z)=
\frac{N}{\pi \sqrt{1-\tau^2}}
\exp \left\{-\frac{N}{1-\tau^2}
\left[|Z|^2 - \frac{\tau}{2}\left(Z^2+{Z^*}^2\right) \right]  \right\}
\sum_{n=1}^{N-1} \frac{\tau^n}{n!}
\left| H_n \left( \sqrt{\frac{N}{\tau}}Z \right)\right|^2
\end{equation}

This exact result is valid for every finite $N$.
The rest of this section is devoted to sampling
of statistical information that can be obtained from
Eqs.\ (\ref{b4a}) -- (\ref{K}) for large matrix dimensions $N\gg 1$.
First we briefly examine the regime of strong non-Hermiticity
when the real and imaginary parts of a typical eigenvalue
are of the same order of magnitude when $N \to \infty $.
This regime is realized when $0\le\lim_{N\to\infty}\tau<1 $
(recall that for $\tau =1$ our
matrices are Hermitian). We will show that in this case  $\tau
$-dependence of the eigenvalue correlations
on the {\em local} scale becomes essentially trivial
and the correlations become identical to those
found by Ginibre in the case of maximum non-Hermiticity ($\tau = 0$).

Then we will examine the regime of weak non-Hermiticity
when the imaginary part of typical eigenvalue is of the
order of the mean separation between the nearest eigenvalues
along the real axis.
This regime is realized when
\begin{equation}\label{wnh}
\tau = 1-\frac{\alpha^2}{2N}, \;\; \alpha > 0 .
\end{equation}
We will show that by varying the parameter $\alpha $ one
can describe the crossover from the Wigner-Dyson eigenvalue statistic
typical for Hermitian random matrices to the Ginibre
eigenvalue statistic typical for non-Hermitian random matrices.

To begin with the regime of strong non-Hermiticity
we first recall that in this regime the eigenvalues in bulk
are confined to an ellipse in the complex plane and they
are distributed there with constant density (cf. Eq.(\ref{girko})) :
\[
\lim_{N\to \infty} \frac{1}{N} R_1(Z)=
\left\{
\begin{array}{ll}
\displaystyle{
\frac{1}{\pi (1-\tau^2)}
             }, & \mbox{if} \;\;
\displaystyle{
\frac{X^2}{(1+\tau)^2}
+ \frac{Y^2}{(1-\tau)^2} \le 1
             }\\
0, & \mbox{otherwise}.
\end{array}
\right.
\]
This fact can be inferred from Eq.\ (\ref{den}). Inside the ellipse
every domain of linear dimension of the order of $1/\sqrt{N}$
contains, on average, a finite number of eigenvalues. Thus
the eigenvalue statistics on the local scale are determined
by the correlation functions
\[
\tilde R_n (z_1, \ldots , z_n)
\equiv N^{-n} R_n (\sqrt{N}Z_1, \ldots , \sqrt{N}Z_n)
\]
of the rescaled eigenvalues $z=\sqrt{N}Z$.
This rescaling is effectively equivalent to
the particular normalization of the distribution
(\ref{P(J)}) which yields $\langle \Tr JJ^{\dagger } \rangle
=N^2$, the normalization used in \cite{Gin}.

One can easily evaluate the rescaled correlation functions
$\tilde R_n$ exploiting Mehler's formula\footnote{Mehler's
formula can be derived
by using the integral representation
(\ref{H}) for Hermite polynomials in the
l.h.s. of Eq.\ (\ref{Mehler}).} \cite{S}:
\begin{equation}\label{Mehler}
\sum_{n=1}^{N-1} \frac{\tau^n}{n!}
H_n \left( \frac{z_1}{\sqrt{\tau}}\right)
H_n \left( \frac{z_2^*}{\sqrt{\tau}}\right)=
\frac{1}{\sqrt{1-\tau^2}}
\exp
\left\{
\frac{1}{1-\tau^2}
       \left[
          z_1z_2^* - \frac{\tau}{2}
             \left(
                 z_1^2+{z_2^*}^2
             \right)
       \right]
\right\}.
\end{equation}
Indeed, denote $\tilde K_N(z_1,z_2)=N^{-1}K_N(\sqrt{N}z_1,
\sqrt{N}z_2)$.
Then, by Mehler's formula
\[
\lim_{N\to \infty}
\tilde K_N(z_1,z_2)=
\frac{1}{\pi (1-\tau^2)}
\exp
\left[
   \frac{2z_1z_2^* -|z_1|^2 - |z_2|^2}{2(1-\tau^2)}
\right]
\exp
\left[
    \frac{\tau ({z_1^*}^2-z_1^2 + z_2^2 -{z_2^*}^2 )}{4(1-\tau^2 )}
\right]
\]
and in view of the relationship
\[
\tilde R_n(z_1,...,z_n)=\det \left[ \tilde K_N (z_j,z_k)
\right]_{j,k=1}^n,
\]
one obtains that
\begin{equation}\label{Gin}
\lim_{N\to \infty} \tilde R_n (z_1, \ldots , z_n)=
\left[
    \frac{1}{ \pi  (1-\tau^2)   }
\right]^n
\exp
\left(
  - \frac{1}{1-\tau^2}\sum_{j=1}^n |z_j|^2
\right)
\det
\left[
   \exp
     \left(
        \frac{z_jz_k^*}{1-\tau^2}
     \right)
\right]_{j,k=1}^n.
\end{equation}
In particular
\begin{eqnarray}
\label{Gin1}
\lim_{N\to \infty} \tilde R_1 (z)&=& \frac{1}{ \pi  (1-\tau^2)   }\\
\label{Gin2}
\lim_{N\to \infty} \tilde R_2 (z_1, z_2)&=&
\left[
   \frac{1}{ \pi  (1-\tau^2)   }
\right]^2
\exp
      \left[
         -\frac{|z_1-z_2|^2}{1-\tau^2}
      \right].
\end{eqnarray}
After the natural additional rescaling
$z/\sqrt{1-\tau^2} \to z $
Eqs.\ (\ref{Gin}) -- (\ref{Gin2}) become identical to those
found by Ginibre \cite{Gin} for the case $\tau =0$.

Now we move on to the regime of weak non-Hermiticity [see
Eq.\ (\ref{wnh})]. We will show that in this regime
new non-trivial correlations occur
on the scale:
$\im Z_{1,2}=O(1/N)$, $\re Z_1 -\re Z_2=O(1/N)$.

To find the density of complex
eigenvalues and describe their correlations,
let us define new variables
$y_1$, $y_2$, $\omega $:
\begin{equation}\label{b8}
Z_1=X+\frac{\omega }{2N}+i\frac{y_1}{N}, \; \;
Z_2=X-\frac{\omega }{2N}+i\frac{y_2}{N}.
\end{equation}
Our first goal is to evaluate the kernel
$K_N (Z_1, Z_2)$ in the limit
\begin{equation}\label{b9}
N\to\infty , N(1-\tau) \to \alpha^2/2,
\; \; X, \omega , y_{1,2}\, \mbox{are}\,\, \mbox{fixed}.
\end{equation}
Using the integral representation for Hermite polynomials,
Eq.\ (\ref{H}), we can rewrite  $K_N (Z_1, Z_2)$ in the form
\begin{eqnarray*}
K_N (Z_1, Z_2)&=&\frac{N}{\pi \sqrt{1-\tau^2}}
\exp
\left[
     \frac{N(Z_1^2+{Z_2^*}^2)}{2\tau }
     -
     \frac{N(|Z_1|^2+|Z_2^*|^2)}{2(1-\tau^2) }
     +
     \frac{N\tau (Z_1^2+Z_2^2+{Z_1^*}^2+{Z_2^*}^2)}{4(1-\tau^2)}
\right]\times \\
 & &
\sum_{n=0}^{N-1}\frac{\tau^n}{n!} \frac{1}{2\pi}
\int\limits_{-\infty}^{+\infty}\! dt\
\int\limits_{-\infty}^{+\infty}\! ds\
(ts)^n \exp
    \left[
        -\frac{t^2+s^2}{2}+\sqrt{\frac{N}{\tau}} \left(
itZ_1-isZ_2^* \right)
    \right]
\end{eqnarray*}
Using new variables (\ref{b8}) in the equation above
and making the substition
\[
u=(t+s)\sqrt{\frac{N}{\tau}} \ \mbox{and}\
v=(t-s)\sqrt{\frac{N}{\tau}}
\]
in the integrals, we obtain
\begin{eqnarray*}
K_N (Z_1, Z_2)&=&
\frac{N^2}{\tau \sqrt{1-\tau^2}}
\exp
\left[
 \frac{\omega^2}{4N\tau (1+\tau )}
 -
 \frac{y_1^2+y_2^2}{2N\tau (1-\tau )}
 +
 \frac{iX(y_1-y_2)}{\tau}
 +
 \frac{i\omega (y_1+y_2)}{2N\tau}
\right]\times \\
& &
\frac{1}{2\pi}
\int\limits_{-\infty}^{+\infty}\! dv\
\exp
\left\{
   \left[
       -\frac{Nv^2}{4}
          \left(
              1+\frac{1}{\tau}
          \right)
       + \frac{iXvN}{\tau} +
       \frac{NX^2}{\tau (1+\tau )}
  \right]
  - \frac{v(y_1-y_2)}{2\tau}
\right\} \times \\
& &
\frac{1}{2\pi}
\int\limits_{-\infty}^{+\infty}\! du\
\exp
\left\{
  -\frac{Nu^2}{4}
    \left(
      \frac{1}{\tau}-1
    \right)  + \frac{iu\omega}{2\tau} -\frac{u(y_1+y_2)}{2\tau}
\right\}
\Theta_N \left[ \frac{N}{4}(u^2-v^2) \right],
\end{eqnarray*}
were we have introduced the notation
\[
\Theta_N (x) =e^{-x} \sum_{n=1}^{N-1} \frac{x^n}{n!}, \;\; x\ge 0.
\]

Now we are in a position to evaluate $K_N(Z_1, Z_2)$ in
the regime defined by Eq.(\ref{b9}).
Indeed, in this regime, to the leading order,
\begin{eqnarray*}
\left[
 -\frac{Nv^2}{4}
    \left(
       1+\frac{1}{\tau}
    \right)
    + \frac{iXvN}{\tau} +
    \frac{NX^2}{\tau (1+\tau )}
\right]
             & =  & -\frac{N(v-iX)}{2}\\
\left\{
  -\frac{Nu^2}{4}
    \left(
      \frac{1}{\tau}-1
    \right)  + \frac{iu\omega}{2\tau} -\frac{u(y_1+y_2)}{2\tau}
\right\}
              & =  &
-\frac{u^2\alpha^2}{8} +\frac{iu\omega}{2} -\frac{u(y_1-y_2)}{2}.
\end{eqnarray*}
>From these relations one obtains that
\begin{eqnarray*}
K_N (Z_1, Z_2)& =  &
\frac{1}{2\pi}
\frac{N^2}{\alpha }
\exp
\left[
   -\frac{y_1^2+y_2^2}{\alpha^2}+iX(y_1-y_2)
\right]\times \\
 & &
\int\limits_{-\infty}^{+\infty}\! \frac{du}{\sqrt{2\pi}}
\exp
\left[
   -\frac{u^2\alpha^2}{8}+\frac{iu\omega}{2} -\frac{u(y_1+y_2)}{2}
\right] \times \\
& &
\frac{N}{\sqrt{2\pi}}
\int\limits_{-\infty}^{+\infty}\! dv\
\exp
\left[
   -\frac{N(v-iX)^2}{2}-\frac{v(y_1-y_2)}{2}
\right]
\Theta_N\left[ \frac{N}{4} (u^2-v^2)\right].
\end{eqnarray*}
Taking into the account that
\begin{equation}\label{b10}
\lim_{N\to \infty } \Theta_N (Nx)=
\left\{
\begin{array}{ll}
1, & 0\le x < 1 \\
0, & x > 1
\end{array}
\right.
\end{equation}
and evaluating
the integral over $v$ by the saddle point method
we finally obtain that in the regime
Eq. (\ref{b9}),
to the leading order,
\begin{eqnarray}\label{kern}
K_N
\left(
X+\frac{\omega}{2N}+\frac{iy_1}{N},
X-\frac{\omega}{2N}+\frac{iy_2}{N}
\right) =
\frac{N^2}{\pi\alpha}
\exp{\left\{- \frac{y_1^2+y_2^2}{\alpha^2}+\frac{iX (y_1-y_2)}{2}
\right\}}g_{\alpha}\left(y-\frac{i\omega}{2}\right),
\end{eqnarray}
where $ y=(y_1+y_2)/2$ and
\begin{equation}\label{g}
g_{\alpha}(y)=\int\limits_{-\pi\nu(X)}^{\pi\nu(X)}\frac{du}{\sqrt{2\pi}}
\exp{\left(-\frac{\alpha^2u^2}{2}-2uy\right)},
\end{equation}
with
$\nu(X)=\frac{1}{2\pi}\sqrt{4-X^2}$ standing for the Wigner
semicircular density of
real eigenvalues of the Hermitian part $\hat{H_1}$ of the matrices
$\hat{J}$.

Equation (\ref{kern})
constitutes the most important result in this section.
The kernel $K_N$ given by Eq.\ (\ref{kern})
determines all the properties of complex eigenvalues in the regime of
weak non-Hermiticity. For instance,
the mean value of the density
$\rho(Z)= \sum_{i=1}^N\delta^{(2)}(Z-Z_i)$
of complex eigenvalues $Z=X+iY$ is  given by $
\langle\rho(Z)\rangle= K_N(Z,Z)$.
Putting $y_1=y_2$ and $\omega=0$ in Eqs.\ (\ref{kern})--(\ref{g})
we immediately recover the density
 Eq.(\ref{13}) found by the supersymmetry approach
\footnote{In the present section
we normalized $\hat{H}_2$ in such a way that for weak
non-Hermiticity regime
we have
$\lim_{N\to\infty} \mbox{Tr}\hat{H}_2^2=N$, whereas
the normalization Eq.(\ref{norm}) gives
$\lim_{N\to\infty} \mbox{Tr}\hat{H}_2^2=N(1+w^2)$.
It is just because of this difference the parameter
$\tilde{a}$ entering Eq.(\ref{13})
contains extra factor $1+w^2$ as compared to the present case.}.

One of the most informative statistical measures of the spectral
correlations
is the `connected' part of the two-point correlation function of
eigenvalue
densities:
\begin{equation}
\label{defy}
\left\langle \rho(Z_1)\rho(Z_2)\right\rangle_c
=\left\langle
\rho(Z_1)\right\rangle \delta^{(2)}(Z_1-Z_2)-{\cal Y}_2(Z_1,Z_2),
\end{equation}

In particular, it determines the variance $\Sigma^2(D)=
\langle n(D)^2\rangle-\langle n(D)\rangle^2$ of the number
$n=\int\limits_D d^2Z\rho(Z)$
of complex eigenvalues in any domain $D$ in the complex plane,
see the Appendix for a detailed exposition.

Comparing with the definitions, Eqs.\ (\ref{R_n}) and (\ref{b4a})
 we see that
the {\it cluster function} ${\cal Y}_2(Z_1,Z_2)$
is expressed in terms of the kernel $K_N$
as ${\cal Y}_2(Z_1,Z_2)=\left|K_N(Z_1,Z_2)\right|^2$.

It is evident that in the limit of weak non-Hermiticity
the kernel $K_N$ in Eq.(\ref{kern})
depends on $X$ only via the semicircular density $\nu(X)$.
Thus, it does not change with $X$ on the local scale comparable with
the mean spacing along the real axis $\Delta\sim 1/N$.

The cluster function $ {\cal Y}_2$ is given by the following explicit expression:
\begin{equation}\label{clexp}
{\cal Y}_2(Z_1,Z_2)=
\frac{N^4}{\pi^2 \alpha^2}
\exp
\left(
          -\frac{2(y_1^2+y_2^2)}{\alpha^2}
\right)
\left|
           \int\limits_{-\pi\nu(X)}^{\pi\nu(X)}\!
           \frac{du}{\sqrt{2\pi}}\
           \exp
             \left[
                  -\frac{\alpha^2 u^2}{2}-u(y_1+y_2)+iu\omega
             \right]
\right|^2
\end{equation}
The parameter
$a=\pi\nu(X)\alpha$ controls the deviation from Hermiticity.
When
$a\gg 1$ the limits of integration in Eq.(\ref{clexp})
can be effectively put to $\pm\infty$ due to the Gaussian cutoff
of the integrand. The corresponding Gaussian
integration is trivially performed
yielding in the original variables $Z_1,Z_2$
the expression equivalent (up to a trivial rescaling) to
that found by Ginibre \cite{Gin}:
${\cal Y}_2(Z_1,Z_2)=(N^2/\pi \alpha^2)^{2}
\exp\{-N^2|Z_1-Z_2|^2/{\alpha}^2\}$.
In the opposite case $a\to 0$ the cluster function tends to GUE form
${\cal Y}_2(\omega,y_1,y_2)=\frac{N^4}{\pi^2}\delta(y_1)
\delta(y_2)\frac{\sin^2{\pi\nu(X)\omega}}{\omega^2}$.

\begin{figure}
\epsfxsize=0.9\hsize
\epsffile{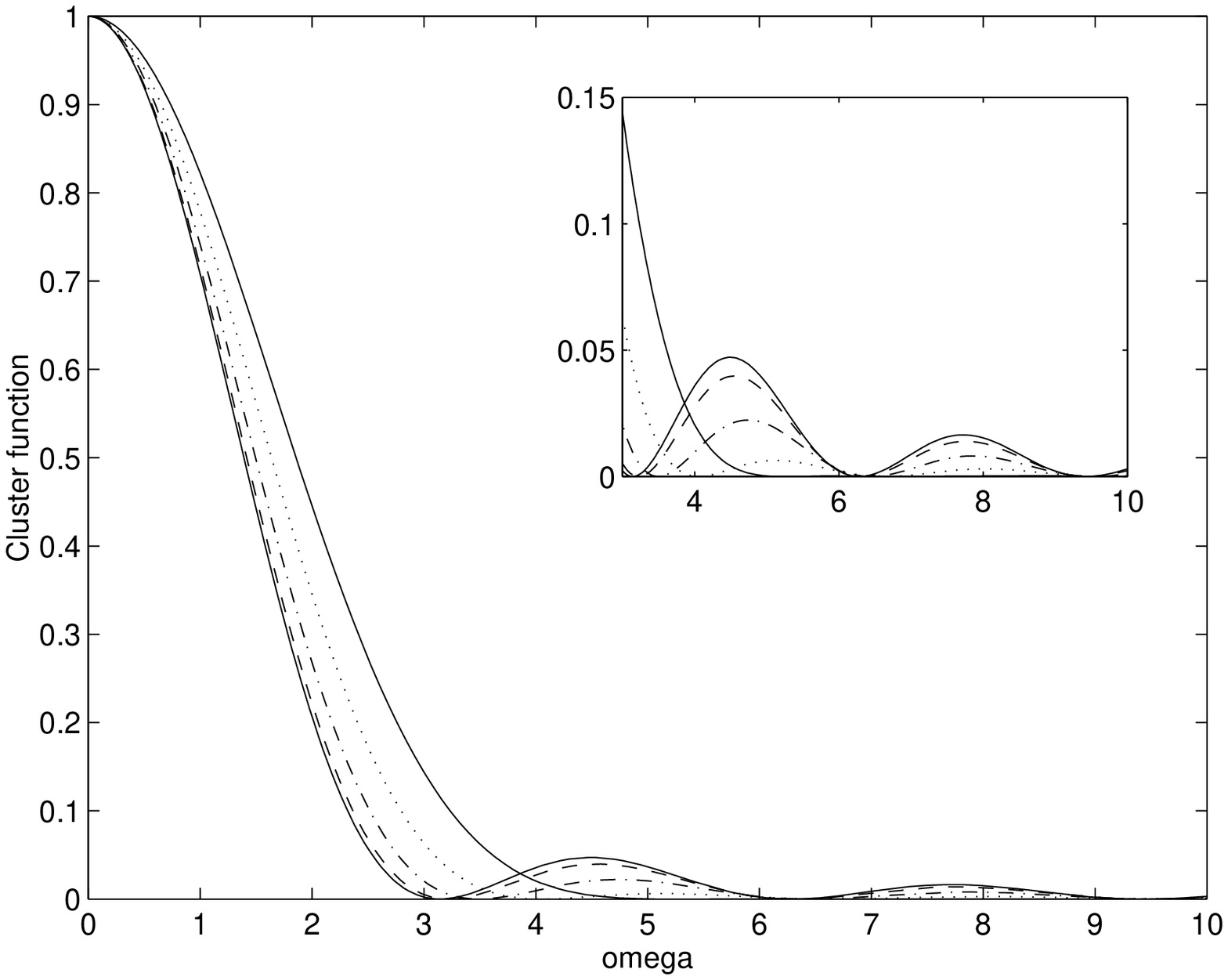}

\vspace{0.1\hsize}
\caption{
\label{figyan1}
The cluster function
$Y_2$,
defined by
Eq.\ (\ref{my_Y2}),
against
$\tilde \omega $
for the parameter values
$\tilde y_1 =\tilde y_2 =0$
and
$a=0$
(left solid line, the GUE case),
$a=0.5$
(dashed line),
$a=1.0$
(dot-dashed line),
$a=1.5$
(dotted line), and
$a=2.0$
(the right solid line, well approximates the Ginibre
case)
}
\end{figure}

One can also define a
 renormalized cluster function:
\[
Y_2(Z_1,Z_2)=\frac{{\cal Y}_2(Z_1,Z_2)}{R_1(Z_1)R_1(Z_2)}.
\]
Introduce the notation
\[
Y_2(\omega , y_1, y_2) \equiv
\lim_{N\to \infty} Y_2
\left(
X+ \frac{\omega}{2N}+\frac{iy_1}{N}, X-\frac{\omega}{2N}+\frac{iy_2}{N}
\right) .
\]
Then
\begin{equation}\label{my_Y2}
Y_2(\omega , y_1, y_2)=
\frac
{
        \displaystyle{
\left|
     \int_0^1\!dt \ \exp
            \left(
                  -\frac{a^2t^2}{2}
            \right)
      \cosh \left[
                t(\tilde y_1+\tilde y_2+i\tilde\omega )
            \right]
\right|^2
                      }
}
{
\displaystyle{
\int_0^1\!dt \ \exp
     \left(
        -\frac{a^2t^2}{2}
     \right)
      \cosh ( 2t\tilde y_1 )
\int_0^1\!dt \ \exp
     \left(
        -\frac{a^2t^2}{2}
     \right)
\cosh ( 2t\tilde y_2 )
           }
},
\end{equation}
where
\[
\tilde y_{1,2}=y_{1,2}\pi \nu (X), \;\;\;\;\;\;\;\;
 \tilde \omega  = \omega\pi \nu (X)\;\;\;\;\;\;\;\;
.
\]
It has advantages of being non-singular in the Hermitian limit $a\to 0$ and coinciding with the usual GUE cluster function
$\tilde \omega^{-2}\sin^2{\tilde \omega}$ on the real axis: $y_1=y_2=0$.
We plotted this function for different values of the parameter $a$ in
Fig.\ \ref{figyan1}.

The operation of calculating the Fourier transform of the cluster
function
over its arguments $\omega,y_1,y_2$ amounts to
 simple Gaussian and exponential integrations. Performing them
one finds
the following expression for the {\it spectral form-factor}:
\begin{eqnarray}\label{for}
b(q_1,q_2,k)&=&\int\limits_{-\infty}^{\infty}\! d\omega \
\int\limits_{-\infty}^{\infty}\! dy_1\
\int\limits_{-\infty}^{\infty}\! dy_2 \
{\cal Y}_2(Z_1,Z_2)\exp [2\pi i(\omega k+y_1q_1+y_2q_2)]\\
\nonumber
 &=& N^4
\exp
\left[
   -\frac{\alpha^2 (q_1^2+q_2^2+2k^2)}{2}
\right]
\frac{\sin{\left[\pi^2 \alpha^2(q_1+q_2)(\nu(X)-|k|)\right]}}
{\pi^2 \alpha^2(q_1+q_2)}\theta(\nu(X)-|k|),
\end{eqnarray}
where $Z_1$ and $Z_2$ are given by Eq.\ (\ref{b8}), and
$\theta(u)=1$  for $u>0$ and zero otherwise.

We see, that everywhere in the regime of weak non-Hermiticity
$0<\alpha<\infty$ the formfactor shows a kink-like behaviour
at $|k|=\nu(X)$. This feature is inherited from the corresponding
Hermitian counterpart-the Gaussian Unitary Ensemble. It
reflects the oscillations of the cluster function with
$\omega$ which is a manifestation of the long-ranged
order in eigenvalue positions along the real axis\cite{Bohigas}.
When non-Hermiticity increases
the oscillations become more and more damped
as is evident from the Fig.\ \ref{figyan1}.

As is well-known\cite{Mehta,Bohigas}, the knowledge of the formfactor
 allows one to determine the variance $\Sigma_2$ of a number of
eigenvalues in any domain $D$ of the complex plane.
Small $\Sigma_2$ is a signature of a tendency for levels to form
a crystal-like structure with long correlations.
In contrast, increase in the number variance signals about growing
decorrelations of eigenvalues.

For the sake of completeness
we derive  the corresponding relation in the Appendix,
see Eq.(\ref{formf1}). In the general case
this expression is not very transparent, however. For this reason we
restrict ourselves to the simplest case,
choosing  the domain $D$ to be the infinite strip of width $L_x$
(in units of mean spacing along the real axis $\Delta=(\nu(0)N)^{-1}$)
oriented perpendicular to the real axis:
$0<\mbox{Re}Z<~L_x\Delta;\quad -\infty
<\mbox{Im}Z<\infty$.
Such a choice means that we look only at real parts of complex
eigenvalues irrespective of their imaginary parts.
It is motivated, in particular, by
the reasons of comparison with the GUE case, for which  the function
$\Sigma_2(L_x)$ behaves at large $L_x$ logarithmically:
$\Sigma_2(L_x)\propto \ln{L_x}$ \cite{Bohigas}.

After simple calculations (see Appendix) one finds
\footnote{In our earlier Letter \cite{FKS2} the expression
Eq.(\ref{var}) and formulae derived from it
 erroneously contained $\pi a$ instead of $a$.}
\begin{equation}
\label{var}
\Sigma_2(L_x)=
L_x
\left\{
    1-2
    \int\limits_0^{L_x}\! dk \
    \left(
          1-\frac{k}{L_x}
     \right)
     \frac{\sin^2 (\pi k)}{(\pi k)^2}
    \exp
    \left[
        -\left(\frac{a k}{L_x}\right)^2
    \right]
\right\}
\end{equation}

\begin{figure}
\epsfxsize=0.9\hsize
\epsffile{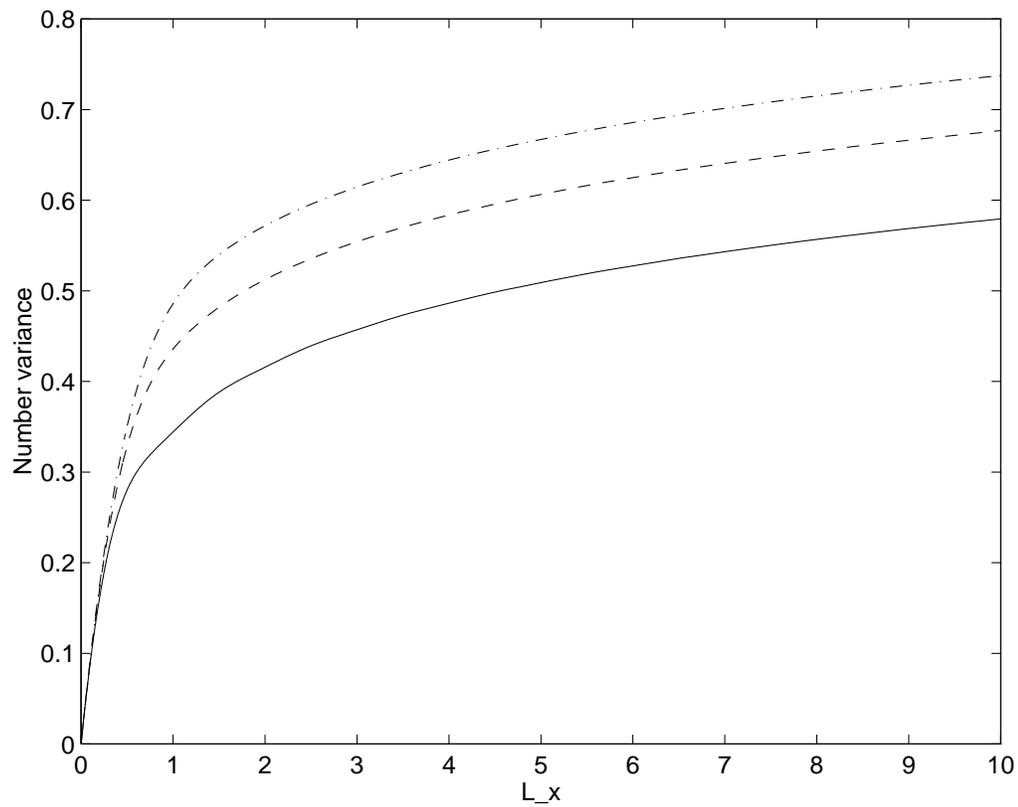}

\vspace{0.1\hsize}
\caption{
\label{figyan2}
The number variance $\Sigma_2$, defined by Eq.\ (\ref{var}),
against $L_x$ for the parameter values
$a=0$ (solid line, the GUE case), $a=1.5$ (dashed line),
and
$a=2.0$ (dot-dashed line)
}
\end{figure}

First of all, it is evident that $\Sigma_2$ grows systematically
with increase in the degree of non-Hermiticity $a=\pi\nu(0)\alpha$,
see Fig.\ \ref{figyan2}.
This fact signals on the gradual decorrelation of the {\it real}
parts $\re Z_i$ of complex eigenvalues. It can be easily understood
because of increasing possibility for eigenvalues to avoid one
another along the $Y=\im Z$ direction, making their projections on the
real axis $X$ to be more independent.

In order to study the difference from
the Hermitian case in more detail let us consider again the large
$L_x$ behaviour.
Then it is evident, that
 the number variance is only slightly modified by non-Hermiticity
as long as $a\ll L_x$. We therefore consider the case
 $a\gg 1$ when we expect essential differences from the Hermitian case.
For doing this it is convenient to rewrite Eq.(\ref{var})
as a sum of three contributions:
\begin{eqnarray}\label{var1}
\Sigma_2(L_x)&=&\Sigma_2^{(1)}+\Sigma_2^{(2)}+
\Sigma_2^{(3)}    \\
\nonumber
\Sigma_2^{(1)}&=& L_x
\left\{
1-\frac{2}{\pi^2}\! \int\limits_0^{\infty}\!
\frac{dk}{k^2}\sin^2{(\pi k)}
\exp \left[-\left( \frac{a k}{L_x}\right)^2 \right]
\right\}\\
\nonumber
\Sigma_2^{(2)}&=&\frac{2}{\pi^2}\! \int\limits_0^{\infty}\!
\frac{dk}{k}\sin^2{(\pi k)}
\exp \left[-\left( \frac{a k}{L_x}\right)^2 \right]\\
\nonumber
\Sigma_2^{(3)}&=&\frac{2}{\pi^2}\! \int\limits_1^{\infty}\!
\frac{dz}{z^2}(1-z)\sin^2{\left(\pi z L_x \right)}
\exp [-(a z)^2 ]
\end{eqnarray}

First of all we notice, that
 for large $a$ the third contribution $\Sigma_2^{(3)}$
is always of the order $O(\exp{-a^2)}$ and
can be neglected.

The relative order of the first and second terms
depends on the ratio $L_x/a$. In a large domain $1\ll L_x\sim a$,
the second term $\Sigma_2^{(2)}$ is much smaller
than $\Sigma_2^{(1)}$. This implies that
the number variance grows like
 $\Sigma(L_x)=L_xf(L_x/a)$. We find it more transparent to rewrite
the function $f(u)$ in an equivalent form:
$$
f(u)=1+\frac{2}{\sqrt{\pi}}\left\{\frac{1}{2\pi
u}\left(1-e^{-\pi^2u^2}\right)-
\int\limits_{0}^{\pi u}dte^{-t^2}\right\}.
$$
which can be obtained from Eq.(\ref{var1}) after a simple
transformation.

For $u=L_x/a\ll 1$ we have simply $f\approx 1$
 and hence a linear growth of the number variance. For $u\gg 1$ we have
$f~\approx~(\pi^{3/2}u)^{-1}$. Thus, $\Sigma_2(L_x)$ slows down:
$\Sigma_2(L_x)\approx \frac{a}{\pi^{3/2}}$.

Only for exponentially large $L_x$ such that
$\ln{(L_x/a)}\gtrsim a$
the term $\Sigma_2^{(2)}$ produces a contribution
comparable with $\Sigma_2^{(1)}$. To make this fact evident
 we rewrite
$\Sigma_2^{(2)}$ as:
\begin{equation}\label{sig3}
\Sigma_2^{(2)}=-\frac{2}{\pi^2}\! \int\limits_0^{\infty}\! dk
\ln{\left(\frac{L_x k}{ a}\right)}\frac{\partial}{\partial k}\left\{
\sin^2{\left(\frac{\pi L_x k}{a}\right)}e^{-k^2}\right\}
\end{equation}
For $L/a\gg 1$ we can neglect the oscillatory part of the integrand
effectively substituting $1/2$ for
$\sin^2{\frac{\pi L_x k}{a}}$ in Eq.(\ref{sig3}). The resulting
 integral can be evaluated explicitly. Remembering
that $\Sigma_2^{(1)}\mid_{L_x>>a}\approx a/(2\pi^{3/2})$ we
finally find:
\[
\Sigma_2(L_x\gg a)= \frac{a}{\pi^{3/2}}+\frac{1}{\pi^2}
\left(\ln{\left(\frac{L_x}{ a}\right)}-\frac{\gamma}{2}\right)
\]
where $\gamma$ is Euler's constant. This logarithmic growth
of the number variance is reminiscent of that
typical for real eigenvalues of the Hermitian matrices.

Another important spectral characteristics which
can be simply expressed in terms of the cluster function is
the small-distance
 behavior of the nearest neighbour distance distribution
\cite{Mehta,Bohigas,Oas}. We present the derivation of
the corresponding relationship in Appendix.

Substituting the expression Eqs.(\ref{13},\ref{clexp}) for the mean
density and the cluster function into Eq.(\ref{small})
 one arrives after
a simple algebra to the probability density to have one
eigenvalue at the point $Z_0=X+iy_0\Delta$ and its
closest neighbour at the distance $|z_1-z_0|=s\Delta,\,\,
\Delta=(\nu(X)N)^{-1}$, such that
$s\ll 1$:
\begin{eqnarray} \label{nns}
\nonumber
p_{\alpha}(X+iy_0\Delta,s\Delta)|_{s\ll 1}&=&
\frac{1}{2\pi}\left[
g_{\alpha}(y_0)\frac{\partial^2}{\partial y_0^2}g_{\alpha}(y_0)-\left(
\frac{\partial}{\partial y_0}g_{\alpha}(y_0)\right)^2\right]
\exp
   \left(
       -4\frac{y_0^2}{a^2}
   \right)
\frac{s^3}{a^2}\times \\
 & &
\int\limits_0^{\pi}d\theta
  \exp
\left[
    -\frac{2}{a^2}(s^2\cos^2{\theta}-2y_0s\cos{\theta})
\right]
\end{eqnarray}
where $g_{\alpha}$ is given by Eq.(\ref{g}).

First of all it is easy to see that in the limit $a\gg 1$
one has: $p_{a\gg 1}(Z_0,s\ll 1)
=\frac{2}{\pi} (s/a^2)^3$ in agreement with the cubic repulsion generic
for strongly non-Hermitian random matrices\cite{Gin,diss,Oas}.
On the other hand one can satisfy oneself that in
 the limit $a\to 0 $ we are back to the
familiar GUE quadratic level repulsion: $p_{a\to 0}(Z_0,s\ll 1)
\propto \delta(y_0) s^2$.
In general, the expression Eq.(\ref{nns})
describes a smooth crossover between the two regimes, although
for any $a\ne 0$ the repulsion is always cubic for $s\to 0$.

To this end, an interesting situation may occur when
 deviations from the Hermiticity are very weak: $a\ll \sqrt{2}$ and
`observation points' $Z_0$ are situated sufficiently far
from the real axis: $2|y_0|/a\gg 2^{-1/2}$.

Under this condition the following three regions for the
parameter $s$ should be distinguished:
i)$\frac{s}{a}\ll \frac{a}{4|y_0|}$
ii)$\frac{a}{4|y_0|}\ll \frac{s}{a}\ll 2\frac{|y_0|}{a}$
and finally iii) $ 2^{-1/2}\ll 2\frac{|y_0|}{a}\ll \frac{s}{a}\ll
a^{-1}$.

In the regimes i) and ii) the term linear in $\cos{\theta}$
in the exponent of Eq.(\ref{nns}) dominates yielding the
result of integration to be the modified Bessel function
$\pi I_0\left(\frac{4y_0s}{a^2}\right)$. In the regime iii)
the term quadratic in $\cos{\theta}$ dominates producing
$2\pi e^{-(s/a)^2}I_0\left[(s/a)^2\right]\approx
\left(2\pi a/s\right)^{1/2}$.
As the result, the distribution
$p(Z_0,s)$ displays the following behaviour:
\begin{eqnarray}
\label{5/2}
p_{\alpha}(Z_0,s)&=&
\left\{
\begin{array}{cl}
   \displaystyle{
   \frac{s^3}{a^2},
                 }& \mbox{ for} \quad
   \displaystyle{
    \frac{s}{a}\ll \frac{a}{4|y_0|}
                 }\\
   \displaystyle{
    \frac{s^{5/2}}{2a\sqrt{2\pi|y_0|}},
                 } &\mbox{ for} \quad
   \displaystyle{
   \frac{a}{4|y_0|}\ll
   \frac{s}{a}\ll 2\frac{|y_0|}{a},
                 }\\
   \displaystyle{
   \sqrt{\frac{2}{\pi}}
   \frac{s^2}{a},
                  }
    &\mbox{ for} \quad
    \displaystyle{
    2\frac{|y_0|}{a}\ll \frac{s}{a}\ll
    a^{-1}
                 }
\end{array}
\right\}
\times \\
\nonumber
 & &
\frac{1}{2}
\left[
     g_{0}(y_0)\frac{\partial^2}{\partial y_0^2}g_{0}(y_0)-
       \left(
              \frac{\partial}{\partial y_0}g_{0}(y_0)
       \right)^2
\right]
\exp \left(
           -\frac{4y_0^2}{a^2}
      \right)
\end{eqnarray}
with $g_0(y)\equiv g_{\alpha}(y)|_{\alpha=0}$.

Unfortunately,  the unusual power law $p(s)\propto s^{5/2}$
might be a very difficult one to detect numerically because of
the low density of complex eigenvalues in
the observation points reflected by the presence of the
Gaussian factor in the expression Eq.(\ref{5/2}).

\section{Conclusion}

In the present paper we addressed the issue of eigenvalue statistics
of large weakly non-Hermitian matrices. The regime of weak
non-Hermiticity is defined as that for which the imaginary
part $Im Z$ of a typical complex eigenvalue is of the same order
as the mean eigenvalue separation $\Delta$ for the Hermitian
counterpart.

Exploiting a mapping to the non-linear $\sigma-$model we were able
to show that there are three different "pure" classes of weakly
non-Hermitian matrices: i) almost Hermitian with complex entries
ii) almost symmetric with real entries and iii) complex symmetric
ones. Within each of these classes the eigenvalue statistics
is {\it universal} in a sense that it is the same irrespective of the
particular distribution of matrix entries up to
an appropriate rescaling. There are also crossover regimes between
all three classes.

Our demonstration of universality was done explicitly for
the density of complex eigenvalues of matrices
with independent entries. Within the non-linear $\sigma-$model formalism
one can easily provide a heuristic proof of such a universality
for higher correlation functions as well as for "rotationally
invariant" matrix ensembles, see \cite{HW}.
The above feature is a great advantage of the supersymmetry technique.

A weak point of that method is a very complicated representation of
the ensuing quantities. It seems, that the explicit evaluation of the
higher correlation functions is beyond our reach at the moment, and
even a calculation of the mean density requires a lot of effort, see
\cite{FKS1,Efnonh}. As a result, at present time
the mean density is known explicitly only
for the cases i) and ii).

Fortunately, because of the mentioned universality
another strategy can be pursued. Namely, one can concentrate
on the particular case of matrices with independent, Gaussian
distributed entries for which alternative analytical techniques might
be available. Such a strategy turned out to be a success for
the simplest case of complex almost-Hermitian matrices, where
we found the problem to be an exactly solvable one by the method of
orthogonal polynomials. This fact allowed us to extract all the
correlation functions in a mathematically rigorous way\cite{FKS2}.

One might hope that combining the supersymmetric method
and the method of orthogonal polynomials one
will be able to elevate our understanding
of properties of almost-Hermitian
random matrices to the level typical for their Hermitian counterparts.

>From this point of view a detailed numerical investigation of
different types of almost-Hermitian random matrices is highly
desirable. Recently, an interesting work in this direction appeared
motivated by the theory of chaotic scattering \cite{reso}.
Weakly non-Hermitian matrices emerging in that theory are
different from the matrices considered in the present paper because
of the
specific form of the skew Hermitian perturbation, see e.g. \cite{FS}.
This fact makes impossible a quantitative comparison of our results
with those
obtained in \cite{reso}. The qualitative fact of increase in number
variance with increase in non-Hermiticity agrees well with our findings.
Let us finally mention
that the knowledge of the {\it time-delay} correlations [\cite{FSR}]
 allows one to make a plausible conjecture about the form of the number variance for the scattering systems with broken time-reversal symmetry.
These results will be published elsewhere [\cite{new}].

 The financial support
       by SFB-237(Y.V.F and H.-J.S.) and EPRSC Research Grant GR/L31913
       (Y.V.F. and B.A.K.) is acknowledged with thanks.
 Y.V.F. is grateful to the School of Mathematical Sciences, Queen Mary\&
Westfield
College, University of London and to the Newton Institute, Cambridge
for the warm hospitality extended to him during
his visits.

\appendix

\section*{Number
variance and nearest neighbour distance distribution.}

The number of eigenvalues in any domain $A$ is expressed in terms
of the eigenvalue density as in Eq.(\ref{b6}).
Then the variance $\Sigma_2(A)=\langle n(A)^2\rangle-\langle
n(A)\rangle^2 $ of the number $n(A)$ is given by:
\begin{eqnarray*}
\Sigma_2(A)&=&
\int\limits_A \! d^2Z_1 \
\int\limits_A  \!d^2Z_2 \
[\langle\rho(Z_1)\rho(Z_2)\rangle-
\langle\rho(Z_1)\rangle\langle\rho(Z_2)\rangle]\\
&=&
\int\limits_A \! d^2Z\
\langle\rho(Z)\rangle-
\int\limits_A \! d^2Z_1\
\int\limits_A \! d^2Z_2\
{\cal Y}_2(Z_1,Z_2),
\end{eqnarray*}
where we used the definition of the cluster function
${\cal Y}_2(Z_1,Z_2)$.

We are interested in finding this variance for the domain $A$ being a
rectangular in the complex plane $Z=X+iY$: $ 0<X<L_x;\quad -L_y<Y<L_y$.
Moreover, we are going to consider the extension $L_x$ being comparable
with the mean eigenvalue separation along the real axis:
$\Delta=1/(\nu(X)N)$.
We know that on such a scale the mean eigenvalue density is independent
of $X$ and can be replaced by its value $\nu(0)$ at $X=0$, whereas
 the cluster function depends on $\Omega=X_1-X_2$
rather than on $X_1$ and $X_2$ separately. Using these facts, we obtain:
\begin{equation}
\Sigma_2(L_x;L_y)=L_x\int\limits_{-L_y}^{L_y}\! dY\
\langle\rho(0,Y)\rangle-
2\int\limits_{0}^{L_x}\!d\omega \
(L_x-\Omega)
\int\limits_{-L_y}^{L_y}\! dY_1 \
\int\limits_{-L_y}^{L_y}\! dY_2 \
{\cal Y}_2(|\Omega|,Y_1,Y_2)
\end{equation}
It is convenient to introduce the {\it spectral form-factor}
$B(K,Q_1,Q_2)$ by the Fourier transform:
\begin{equation}\label{formf}
 {\cal Y}_2(|\Omega|,Y_1,Y_2)=
\int\limits_{-\infty}^{\infty}\! dK\
\int\limits_{-\infty}^{\infty}\! dQ_1\
\int\limits_{-\infty}^{\infty}\! dQ_2\
B(K,Q_1,Q_2)e^{-2\pi i(\omega K+Y_1Q_1+Y_2Q_2)}
\end{equation}
The number variance can be expressed in
 terms of the spectral form-factor, Eq.(\ref{formf}), as:
\begin{eqnarray}
\label{formf1}
\Sigma(L_x;L_y)&=& L_x\int\limits_{-L_y}^{L_y}\! dY\
\langle\rho(0,Y)\rangle - \\
\nonumber
 & &
\frac{2}{\pi^4}
\int\limits_{-\infty}^{\infty}\! dQ_1\
\int\limits_{-\infty}^{\infty}\! dQ_2\
\frac{\sin{2\pi Q_1}\sin{2\pi Q_2}}{Q_1Q_2}
\int\limits_{0}^{\infty}\! dK\
\frac{\sin^2{\pi K L_x}}{K^2}
B(K,Q_1/L_y,Q_2/L_y).
\end{eqnarray}

In particular, for the strip $0<X<L; \,\,-\infty<Y<\infty$
the number variance is given by a rather simple expression:
\begin{equation}\label{simp}
\Sigma(L;\infty) =LN\nu(0)-
\frac{2}{\pi^2}
\int\limits_{0}^{\infty}\! dK\
\frac{\sin^2{\pi K L}}{K^2}
B(K,0,0)
\end{equation}

In the main text of the paper we use the variables $y_{1,2}=NY_{1,2}$
and $\omega=N\Omega$. Correspondingly, the form-factor $B(K,Q_1,Q_2)$
is related to $b(k,q_1,q_2)$, Eq.(\ref{for}) as
$$
B(K,0,0)=\frac{1}{N^3}b(K/N,0,0)=
N e^{-\pi^2\alpha^2(K/N)^2}\left(\nu(0)-|K/N|\right)
\theta\left(\nu(0)-|K/N|\right)
$$.

Substituting this expression into Eq.(\ref{simp}) and measuring the
length
$L$ in the units of $\Delta=1/(N\nu(0))$ as $L=\Delta L_x$ we find
after simple
manipulations the eq.(\ref{var}).

Let us now derive the relation between the cluster function and
the nearest neighbour distance distribution $p(Z_0,S)$, see also
\cite{Oas}.

We define the quantity $p(Z_0,S)$ as
 the probability density of the following
event: i) There is exactly one eigenvalue at the point $Z=Z_0$
of the complex plane. ii) Simultaneously, there is exactly one
eigenvalue on the circumference of the circle $|Z-Z_0|=S$
iii) All other eigenvalues $Z_i$ are {\it out} of that circle:
$|Z_i-Z_0|>S$.

As a consequence, the normalization condition is: $\int
d^2Z_0\int\limits_0^{\infty}
dS\, p(Z_0,S)=1$. In particular, for Hermitian matrices with real
eigenvalues one has the relation: $p(Z_0,S)=\delta(\mbox{Im} Z_0)\nu(X_0)
\tilde{p}_X(S)$, with $\tilde{p}_X(S)$ being the conventional
"nearest neighbour spacing" distribution at the
point $X$ of the real axis\cite{Mehta}.

Using the above definition one easily finds the relation:
$p(Z_0,S)=-\frac{d}{dS}H(Z_0,S)$, where $H(Z_0,S)$ has the meaning of
the probability density to have one eigenvalue at $Z=Z_0$ and no
other eigenvalues inside the disk $D$:$|Z-Z_0|\le S$. The latter quantity
is related to the joint probability density of complex eigenvalues
as
\begin{equation}\label{hole}
H(Z_0,S)=\frac{1}{N}\sum_{i=1}^N\int d^2Z_1...\int d^2Z_N{\cal P}_N
(Z_1,...,Z_N)\delta^{(2)}(Z_i-Z_0)\prod_{j\ne
i}\left[1-\chi_D(Z_j)\right]
\end{equation}
where $\chi_D(Z_j)$ is the characteristic function of the disk equal
to unity for points $Z_j$ inside the disc and zero otherwise.

We are interested in finding the leading small-$S$ behaviour
for the function $p(Z_0,S)$. For this one
expands
$$\prod_{j\ne i}\left[1-\chi_D(Z_j)\right]=
1-\sum_{i\ne j}\chi_D(Z_j)+\sum_{j\ne i}\sum_{k\ne i}
\chi_D(Z_j)\chi_D(Z_k)-...
$$
and notices that each factor
$\chi_D(Z_j)$ produces upon integration extra factor
proportional to the area of the disc. Therefore, to the lowest
nontrivial order in $S$ one can restrict oneself by the first two terms
in the expansion and write:
$$H(Z_0,S)\approx\frac{1}{N}\left[\langle\rho(Z)\rho(Z_0)\rangle-
\int d^2Z\chi_D(Z)R_2(Z_0,Z)\right]$$
where we used the definitions of the mean eigenvalue density and
the spectral correlation function, see Eq.(\ref{R_n}).
At last, exploiting that $\int d^2Z\chi_DF(Z)=\int\limits_0^S dr
r\int\limits_0^{2\pi}d\theta F\left(Z_0+re^{i\theta}\right)$ one finally
finds after
differentiation over $S$:
\begin{equation}\label{small}
p(Z_0,S)\approx \frac{S}{N}\int\limits_0^{2\pi}d\theta
\left[\langle\rho(Z_0)\rangle\langle\rho
\left(Z_0+Se^{i\theta}\right)\rangle-
{\cal Y}_2\left(Z_0,Z_0+Se^{i\theta}\right)\right]\end{equation}
where we used the definition of the cluster function, Eq.(\ref{defy}).

In the regime of weak non-Hermiticity this
formula is valid as long as the parameter $S$ is small
in comparison with a typical separation between real eigenvalues
of the Hermitian counterpart: $S\ll \Delta\sim 1/N$.

\newpage

\newpage

\end{document}